\documentclass[sigplan,screen]{acmart}
\usepackage[english]{babel}
\usepackage{subcaption}
\usepackage{amsmath}    % need for subequations
\usepackage{graphicx}   % need for figures
\usepackage{array}
\usepackage{verbatim}   % useful for program listings
\usepackage{tikz}
\usepackage{local-macros}
\usepackage[normalem]{ulem}
\usepackage{textcomp}
\usepackage{float}
\usepackage{placeins}
\usepackage{enumitem}
\usepackage{algpseudocode}
\usepackage{algorithm}
\usepackage{multirow}
\usepackage{xkeyval}
\usepackage{hyperref}
\usepackage{hyperxmp}

\usepackage{pst-poker}

\newcommand*\circled[1]{\tikz[baseline=(char.base)]{
\node[shape=circle,draw,fill=black,inner sep=0.2pt] (char) {{\color{white}#1}};}}

\AtBeginDocument{%  
}

\copyrightyear{2025}
\acmYear{2025}
\setcopyright{cc}
\acmConference[ASPLOS '25]{Proceedings of the 30th ACM International Conference on Architectural Support for Programming Languages and Operating Systems, Volume 1}{March 30--April 3, 2025}{Rotterdam, Netherlands}
\acmBooktitle{Proceedings of the 30th ACM International Conference on Architectural Support for Programming Languages and Operating Systems, Volume 1 (ASPLOS '25), March 30--April 3, 2025, Rotterdam, Netherlands}
\acmDOI{10.1145/3669940.3707221}
\acmISBN{979-8-4007-0698-1/25/03}

\begin{document}


%\ccsdesc[500]{Hardware~Networking hardware}


\title{\sys: An Ultra-Low Latency Ethernet Fabric for Memory Disaggregation}

\author{Weigao Su}
\orcid{0009-0007-0023-798X}
\affiliation{%
\institution{Purdue University}%
\city{West Lafayette, IN}
\country{USA}%
}

\author{Vishal Shrivastav}
\orcid{0000-0003-2770-4799}
\affiliation{%
\institution{Purdue University}%
\city{West Lafayette, IN}
\country{USA}%
}

\begin{abstract}
    % Achieving low remote memory access latency remains the biggest challenge in realizing memory disaggregation over Ethernet within the datacenter. We present EDM that attempts to overcome this challenge using two key ideas. First, while the existing network protocols for remote memory access over Ethernet, such as TCP/IP and RDMA, are implemented on top of Ethernet's MAC layer, EDM takes a rather radical approach of implementing the entire network protocol stack for remote memory access within the Physical layer (PHY) of the Ethernet. This overcomes fundamental latency and bandwidth overheads imposed by the MAC layer, especially for small memory messages. Second, EDM implements a centralized, fast, in-network scheduler for memory traffic within the PHY of the Ethernet switch. Inspired by the classic Parallel Iterative Matching (PIM) algorithm, the scheduler dynamically reserves bandwidth between compute and memory nodes by creating virtual circuits in the switch's PHY, thus eliminating the queuing delay and layer 2 packet processing delay at the switch for memory traffic, with high bandwidth utilization. Our FPGA testbed shows that EDM's network fabric incurs a latency of only $\sim$300 ns for remote memory access in an unloaded network, which is an order of magnitude lower than state-of-the-art Ethernet-based solutions such as RoCEv2 and comparable to emerging PCIe-based solutions such as CXL. Larger-scale network simulations show that even at high network loads, EDM's average latency is within 1.3$\times$ its unloaded latency.

Achieving low remote memory access latency remains the primary challenge in realizing memory disaggregation over Ethernet within the datacenters. We present EDM that attempts to overcome this challenge using two key ideas. First, while existing network protocols for remote memory access over the Ethernet, such as TCP/IP and RDMA, are implemented on top of the MAC layer, EDM takes a radical approach by implementing the entire network protocol stack for remote memory access within the Physical layer (PHY) of the Ethernet.  This overcomes fundamental latency and bandwidth overheads imposed by the MAC layer, especially for small memory messages. Second, EDM implements a centralized, fast, in-network scheduler for memory traffic within the PHY of the Ethernet switch. Inspired by the classic Parallel Iterative Matching (PIM) algorithm, the scheduler dynamically reserves bandwidth between compute and memory nodes by creating virtual circuits in the PHY, thus eliminating queuing delay and layer 2 packet processing delay at the switch for memory traffic, while maintaining high bandwidth utilization. Our FPGA testbed demonstrates that EDM’s network fabric incurs a latency of only $\sim$300 ns for remote memory access in an unloaded network, which is an order of magnitude lower than state-of-the-art Ethernet-based solutions such as RoCEv2 and comparable to emerging PCIe-based solutions such as CXL. Larger-scale network simulations indicate that even at high network loads, EDM’s average latency remains within 1.3$\times$ its unloaded latency.
\end{abstract}
\maketitle

\vspace{-0.25cm}
\section{Introduction}
\label{sec:intro}

Memory disaggregation is a computing architecture in which compute and memory are physically separate nodes of resources connected via a network fabric such as the Ethernet.
Memory disaggregation promises high compute density, fine-grained memory pooling and provisioning, and elastic memory scaling. Hence, it is not surprising that memory disaggregation has attracted significant interest in recent years, both in industry~\cite{hp-the-machine,intelrackscale} and in academia~\cite{osdi16:disaggregated,shoal,legos,mind}.

As the bandwidth of the Ethernet link within the datacenters has reached 100--400 Gbps, with Tbps Ethernet on the horizon~\cite{terabit-eth}, prior works~\cite{osdi16:disaggregated,mind} have shown that the current Ethernet bandwidth within the datacenters is more than sufficient to carry both the memory and traditional IP and storage traffic with minimal bandwidth contention.
However, despite the promise of abundant bandwidth, remote memory access latency remains the key bottleneck in the realization of memory disaggregation over Ethernet. As a result, there have been several proposals for alternative fabrics for memory disaggregation within the datacenters~\cite{cxl,novakovic14sonuma,infiniband}, most recently the PCIe-based CXL fabric~\cite{cxl}, which promise ultra-low latency for remote memory access. However, such fabrics scale poorly in terms of bandwidth, distance, cost, and power compared to Ethernet (\autoref{sec:background:cxl}). Furthermore, even their latency scales poorly with network load (\autoref{sec:eval:simulations}) due to ineffective mechanisms to handle fabric congestion~\cite{aurelia}. 
Hence, an ultra-low latency Ethernet fabric, if realized in practice, has the potential to enable scalable, high bandwidth, and low latency memory disaggregation at low cost and power.

\begin{figure}[t!]
    \centering
    \includegraphics[width=0.48\textwidth]{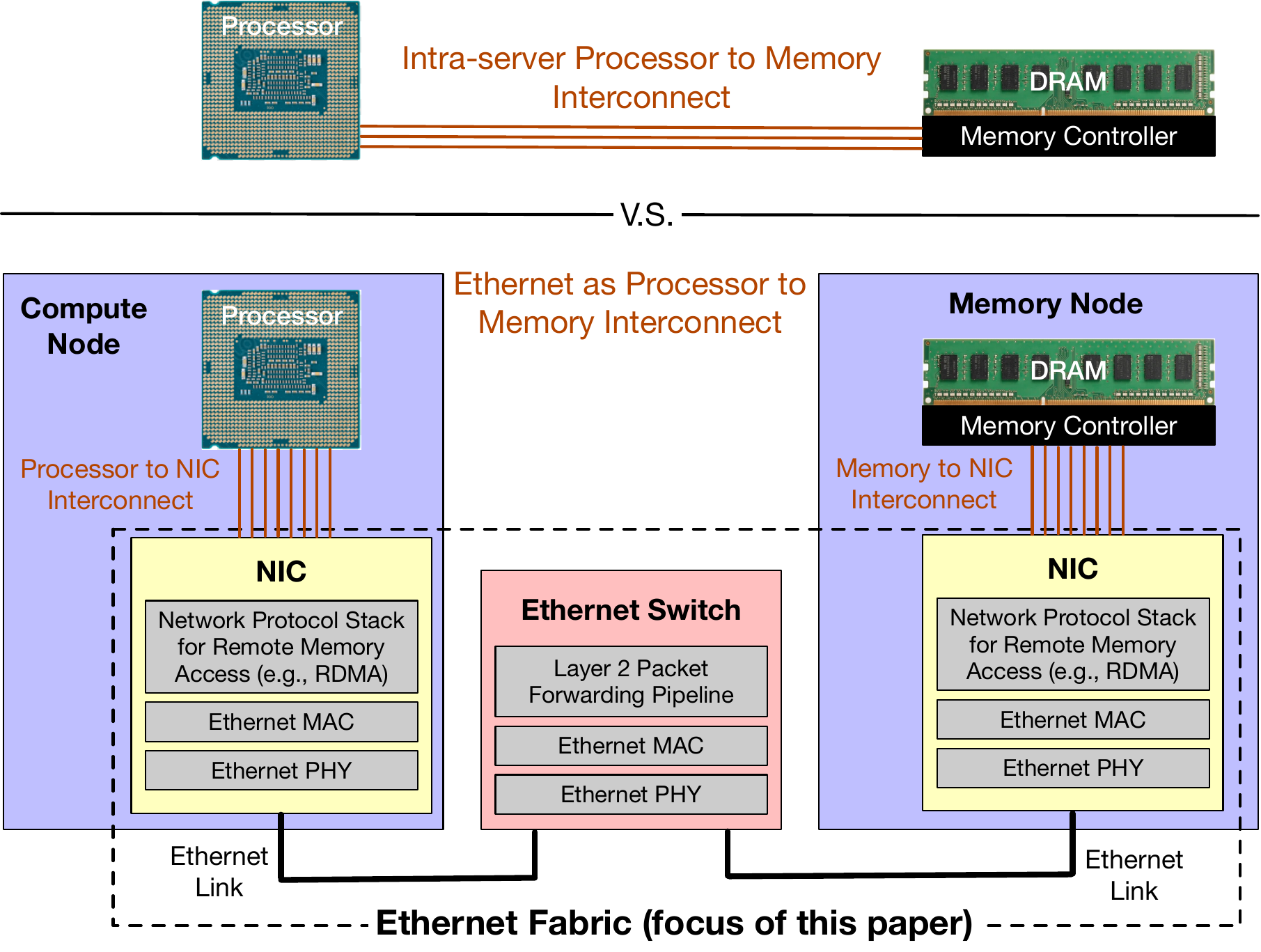}
    \caption{Memory disaggregation over Ethernet.}
    \label{fig:arch}
    \vspace{-0.25cm}
\end{figure}

Using Ethernet as the interconnect between the processor and the memory introduces two sources of latency for memory access (\autoref{fig:arch})---\textit{(i)} the processor/memory to NIC interconnect latency, and \textit{(ii)} the \textbf{Ethernet fabric latency}, which includes the latency of the host network protocol stack for remote memory access over Ethernet (e.g. TCP/IP, RDMA) and the latency of the Ethernet switching network. 

Using the traditional PCIe interconnect between the NIC and the processor/memory would add a latency of the order of few $\mu$s~\cite{pcie_perf}, which is an order of magnitude higher than intra-server memory access latency. Fortunately, in recent years, various new interconnect specifications have been proposed that replace PCIe or build on the physical layer of PCIe, which dramatically reduce the interconnect latency to the NIC~\cite{cxl,ccix,nvlink,opencapi,ucie}. %Most notably, PCIe-based Compute Express Link (CXL)~\cite{cxl} point-to-point peripheral interconnect latency can be of the order of 100\unit{ns}~\cite{li2023pond}. 
The demand for lower latency for cloud services has also prompted tighter integration of the processor and memory with the network controller, which promises to reduce the processor/memory to NIC latency to sub-100\unit{ns}. Such designs exist both in academia (e.g., nanoPU~\cite{nanopu}, soNUMA~\cite{novakovic14sonuma}, FAME-1 RISC-V RocketChip SoC~\cite{firesim}) and industry (e.g., Intel UPI~\cite{upi}). Further, cloud providers are increasingly offloading processing from CPUs to accelerators, such as FPGAs, for cost efficiency and higher performance~\cite{microsoft-fpga,gft,disagg-nf}. Such accelerators~\cite{stratix10,stratixv,alveo} have processor, memory, and network controller integrated on the same motherboard, thus reducing the processor/memory to NIC latency to as few as 10s of nanoseconds (\autoref{sec:eval:prototype}). 

In light of the above (and ongoing~\cite{cc-nic}) architectural and system optimizations to reduce the processor/memory to NIC latency, the Ethernet fabric becomes the primary source of latency for remote memory access. Intra-server memory access latency typically varies from a few 10s to a few 100s of nanoseconds~\cite{dram-latency-2,dram-latency-1}, depending on factors such as the memory access pattern and the location of the memory relative to the processor. However, in existing Ethernet-based networks, even with just a single Ethernet switch and highly optimized host network protocol stacks for remote memory access (e.g., RDMA over Converged Ethernet (RoCEv2)~\cite{rocev2}), the Ethernet fabric latency can be as high as a few $\mu$s~\cite{osdi16:disaggregated} in an unloaded network, which is already an order of magnitude higher than the intra-server memory access latency. And at higher network loads, the fabric latency may easily be several orders of magnitude higher than the unloaded latency due to network queuing and congestion~\cite{shoal}.

Given this, we present \textbf{\sys} (\textbf{\underline{E}}thernet \textbf{\underline{D}}isaggregated \textbf{\underline{M}}emory), which is an ultra-low latency Ethernet fabric for memory disaggregation. Similarly to previous works~\cite{mind,osdi16:disaggregated}, \sys targets rack- or cluster-scale memory disaggregation within the datacenters, where hundreds of compute and memory nodes are connected using a \textit{single} Ethernet switch. To achieve ultra-low latency with high bandwidth utilization, \sys uses two key design ideas, as discussed below.

First, we note that existing Ethernet-based network protocol stacks for remote memory access, such as TCP/IP and RDMA, are implemented on top of Ethernet's Media Access Control (MAC) layer. Unfortunately, the MAC layer imposes fundamental latency and bandwidth overheads for small memory messages due to its operation at frame granularity. In particular, the MAC layer imposes a minimum frame size of 64\unit{B} and an inter-frame gap (IFG) of at least 12 bytes. This results in high bandwidth overhead for small memory messages, such as a remote memory read request that only contains the control information for reading from remote memory, e.g., a 64-bit (8 B)
remote memory address (\autoref{sec:background:sota}). Further, MAC layer does not allow intra-frame preemption. As a result, a small, latency-sensitive memory message cannot preempt the transmission of a large Ethernet frame carrying IP or storage data, thus significantly increasing the latency of small memory messages (\autoref{sec:background:sota}). 

To overcome these challenges, \sys employs a rather radical approach of completely bypassing the MAC layer for remote memory access and implementing the entire network protocol stack for remote memory access inside the Physical Coding Sublayer (PCS) of Ethernet's Physical Layer (PHY), at both the host (\autoref{sec:design:stack:host}) and the switch (\autoref{sec:design:stack:switch}). Note that \sys does not replace the standard Ethernet network stack, which can still be used to carry IP and storage traffic, but rather runs in parallel with it. The key insight \sys uses is that in the 10/25/40/100+ GbE standard, PCS operates at the granularity of 66-bit blocks. This can be leveraged to both reduce the bandwidth overhead for small memory messages as well as enable fine-grained multiplexing of memory messages and non-memory Ethernet frames for lower latency for memory messages. In addition, IFG is also accessible inside PCS, which can be repurposed to carry memory messages, thus further improving bandwidth utilization. %The applications (on the compute node) and the memory controller (on the memory node) sending/receiving memory messages interface directly with \sys in the PHY.
On the TX side, \sys transmits the memory messages using custom 66-bit PHY block types, and on the RX side, \sys extracts the memory messages from the received PHY blocks and replaces them with standard Ethernet PHY block types before sending them to the higher layers. Furthermore, \sys is the first system to enable intra-frame preemption (\autoref{sec:design:stack:isolation}), by multiplexing the transmission of memory messages and non-memory Ethernet frames at the granularity of 66-bit PHY blocks to achieve low latency for small memory messages.%, while ensuring the higher sublayers in PCS at the receiver still receive the PHY blocks of an Ethernet frame in contiguous clock cycles as expected by the standard Ethernet implementation. %\sys ensures this by appropriately buffering the PHY blocks at the receiver until the frame termination block arrives, before releasing the blocks in contiguous cycles to the higher sublayers. The buffering requirement for this design is also minimal, requiring a per-receiver buffer of size at most the maximum allowed Ethernet frame size.

The second key idea in \sys is to implement a centralized memory traffic scheduler in the PHY of the Ethernet switch (\autoref{sec:design:fabric:scheduler}). The scheduler takes as input the current memory traffic demand matrix and implements the classic Parallel Iterative Matching (PIM)~\cite{pim} to dynamically reserve bandwidth between compute and memory nodes by creating virtual circuits in the PHY of the switch. This proactively ensures no queuing and layer 2 packet processing delay at the switch for memory traffic, while guaranteeing high bandwidth utilization. Furthermore, to achieve near-optimal latency under bandwidth contention, \sys augments PIM with priority-based scheduling, such as SRPT~\cite{srtf} scheduling policy.

Implementing \sys's scheduler in practice requires overcoming two key challenges. The first challenge is to acquire the traffic demand matrix for memory traffic accurately and with low overhead. For this, \sys leverages the unique request-reply nature of remote memory reads, where the read request, containing the number of bytes to be read as required by the memory controller interface, e.g., DDR4, implicitly provides a perfectly accurate demand estimate for the corresponding read reply message. Furthermore, by implementing the scheduler on the switch, \sys is able to intercept the read request messages and extract the read demands inline without any bandwidth or latency overhead. In contrast, for remote writes, which are one-sided, \sys does incur the bandwidth and latency overhead of sending explicit demand messages to the switch before sending the actual write message. However, \sys manages to keep the bandwidth overhead low using techniques such as batching, and the latency overhead of RTT/2 is nominal at rack- or cluster-scale. The second challenge is to design a scheduler that can schedule at line rate with small scheduling latency. Na\"{\i}vely, each iteration of the priority-based PIM would take $O(log(n))$ clock cycles to choose the highest priority matching request from a set of $n$ requests for each source port in parallel. In contrast, \sys completes each iteration in a constant number of clock cycles, owing to \sys's novel hardware design (\autoref{sec:design:fabric:hardware}), which intelligently trades off hardware resources for time, by using a combination of a constant-time ordered list data structure, that allows for parallel reads, comparisons, and writes, and a fast priority encoder.

%to select the highest priority demand source to match to each destination port in parallel. Further, due to parallel matching in PIM, a demand from the same source may match to multiple destination ports, in which case, na\"{\i}vely it would take another $O(log(n))$ cycles to resolve the conflict in favor of the highest priority demand source. In contrast, \sys is able to compute this in constant number of clock cycles, owing to \sys's novel hardware design (\autoref{sec:design:fabric:hardware}), that intelligently trades-off hardware resource for time, by using a combination of constant time ordered list data structure that allows for parallel reads, comparisons, and writes, and a fast priority encoder.

We implement \sys's design on FPGAs by extending the PHY of standard 25 Gb Ethernet (\autoref{sec:eval:prototype}). Using a small network hardware testbed of \sys-capable FPGA-based switch and NICs, we show that \sys only incurs a latency of $\sim$300\unit{ns} in an unloaded network, for both the remote memory reads and writes (\autoref{sec:eval:testbed:latency}). The read (write) latency is 3.7$\times$ (1.9$\times$), 6.8$\times$ (3.4$\times$), and 12.7$\times$ (6.4$\times$) lower than the latency of raw Ethernet (standard Ethernet MAC + PHY only), RDMA over Converged Ethernet (RoCEv2), and hardware offloaded TCP/IP network stacks, respectively. Furthermore, \sys's unloaded latency is comparable to both an intra-server two hop NUMA \cite{dram-latency-1} and an unloaded PCIe-based CXL fabric with a single switch hop~\cite{li2023pond}. %(while being more scalable and cost-efficient than CXL (\autoref{sec:background:cxl})). 
Using larger-scale software network simulations in C, we show that even at high network loads, \sys's average latency is within 1.3$\times$ its unloaded latency. Further, using a wide variety of disaggregated network traffic traces obtained from real applications in a loaded network, we show that average message completion time for memory messages in \sys is within 1.2--1.4$\times$ the ideal, and up to 8$\times$ lower than CXL (\autoref{sec:eval:simulations}), whose underlying flow control fails to handle fabric congestion effectively at higher network loads~\cite{aurelia}. %Hence, we argue \sys should be the fabric for rack-scale disaggregation, whereas CXL-based systems might serve as a complementary intra-server memory pool for their lightweight and low-latency advantages.

\section{Background and Motivation}
\label{sec:background}

%In this section, we start by contrasting Ethernet fabric against CXL-based PCIe fabric for memory disaggregation (\autoref{sec:background:cxl}). Next, we discuss the disaggregation model assumed in this paper (\autoref{sec:background:arch}) followed by a discussion on the network memory traffic and its key characteristics that distinguish it from traditional datacenter network traffic (\autoref{sec:background:mem_flow}). Finally, we discuss the limitations of existing Ethernet-based datacenter network stacks when it comes to supporting memory traffic with low latency and high bandwidth utilization (\autoref{sec:background:sota}). %Finally, we discuss the key insights underpinning \sys's design (\autoref{sec:background:edm}).
This section details the background and motivation for \sys.
\vspace{-1cm}
\subsection{Disaggregation Model}
\label{sec:background:arch}
\sys assumes the same disaggregation model as used in previous works~\cite{osdi16:disaggregated,mind}, where separate nodes of resources (compute, memory, storage nodes) are connected via a network fabric. \sys targets rack- or cluster-scale with 100s of nodes connected using a \textit{single} Ethernet switch. State-of-the-art Ethernet switches with 51.2\unit{Tbps}~\cite{512tbps} aggregate capacity could connect up to 512 nodes at 100\unit{Gbps} link speed.
Further, as explained in \autoref{sec:intro}, \sys assumes a low latency processor/memory to NIC interconnect at each compute and memory node, thus making the Ethernet fabric, that includes the host network protocol stack for remote memory access and the Ethernet switching network, as the primary source of latency for remote memory access.

\vspace{-0.5cm}
\subsection{Ethernet vs. Other Disaggregation Fabrics}
\label{sec:background:cxl}

This paper focuses on memory disaggregation over Ethernet.
Two alternate fabrics used in practice for memory disaggregation are Infiniband~\cite{infiniband} and the emerging PCIe-based CXL fabric~\cite{cxl}. However,
 Ethernet-based memory disaggregation has several fundamental advantages over these alternative solutions, especially in the context of disaggregation within datacenters.
%an Ethernet-based fabric, such as RDMA over converged Ethernet (RoCEv2)~\cite{rocev2}, still remains the status quo for memory disaggregation inside datacenters~\cite{osdi16:disaggregated,gu2017efficient,mind,chen2023cowbird}. Following are some of the reasons behind this: 
\textbf{(i) Cost and management.} Ethernet is the default network fabric for traditional datacenter traffic (e.g., IP, storage). Thus, having a separate fabric for memory traffic would result in higher infrastructure and management cost. This is in part the reason why Infiniband has remained limited to the HPC environment. And while CXL is still in its infancy, its cost effectiveness with regard to supporting memory disaggregation within datacenters is already being questioned~\cite{hotnets-cxl}. \textbf{(ii) Bandwidth scaling.} The bandwidth of a CXL switch scales poorly compared to an Ethernet switch, partly due to the limitations of PCIe SerDes, which has much more stringent bit error rate requirements and lower bandwidth per unit of chip area compared to Ethernet SerDes~\cite{semianalysis-cxl}. E.g., state-of-the-art CXL switch supports 16\unit{Tbps} aggregate bandwidth~\cite{cxl-sw,li2023pond} in contrast to 51.2\unit{Tbps} Ethernet switch~\cite{512tbps}. There have also been proposals~\cite{cxl} to use multi-switch CXL fabric for bandwidth scaling, but each switching hop adds around 100 ns of additional latency~\cite{li2023pond} while also making it more challenging to handle fabric congestion. \textbf{(iii) Distance scaling.} CXL-based systems also scale poorly with distance, requiring retimers every 500\unit{mm}~\cite{li2023pond} to preserve signal integrity. In contrast, (optical) Ethernet links can easily scale to several meters, if not more. \textbf{(iv) Fabric congestion.} Both Infiniband and CXL use link level, credit-based flow control to avoid buffer overflow inside the switches. However, due to lack of co-ordination, they are prone to queuing and bandwidth under-utilization, resulting in higher latency and completion time for memory messages~\cite{aurelia}. We evaluate this in \autoref{sec:eval:simulations}.

%Overall, the above reasons highlight the value of building a memory disaggregation system on top of an Ethernet fabric.

\subsection{Memory Traffic in \sys}
\label{sec:background:mem_flow}
\sys abstracts remote memory traffic as either a remote memory \textit{request} message (generated on a compute node) or a \textit{response} message (generated by a remote memory controller). \sys has four remote memory request and response message types as described below. %\sys has four kinds of memory request/response as described below:
\begin{enumerate}[leftmargin=*]
\setlength\itemsep{0.1em}
\item \textbf{Read request (RREQ).} A remote memory read request message generated on a compute node, containing the remote memory address and number of bytes to be read.
\item \textbf{Write request (WREQ).} A remote memory write request message generated on a compute node, containing the remote memory address, number of bytes to be written and the data to be written.
\item \textbf{Read-Modify-Write request (RMWREQ).} An atomic read-modify-write request message for a remote memory location, generated on a compute node. The message contains the remote memory address, an opcode for the modify operation, and arguments for the opcode. \sys implements atomic read-modify-write operations on the NIC at the memory node (\autoref{sec:design:stack:host}), and applications at the compute node could choose one of those operations using the opcode field in the RMWREQ message.
\item \textbf{Read response (RRES).} A message generated by a memory node in response to an RREQ or RMWREQ. For RREQ, the response contains the bytes read from the memory address specified in RREQ. For RMWREQ, the response contains the result of the read-modify-write operation.

\end{enumerate}

A key characteristic of memory traffic that distinguishes it from traditional datacenter network traffic is  extremely \textbf{small message sizes}; at times even smaller than a minimum-sized Ethernet frame (64\unit{B}). In particular, RREQ are inherently small, as they only contain the control information for reading from remote memory, e.g., a 64-bit (8\unit{B}) remote memory address. Similarly, RMWREQ and the corresponding RRES are also inherently small. For example, a popular RMWREQ, namely compare-and-swap~\cite{cas}, contains three 64-bit arguments (24\unit{B}), and its corresponding RRES can be as small as 1 bit True or False. Further, given that a DRAM is byte-addressable with typical word size of 64 bits (in DDR4), even a WREQ and an RRES (in response to an RREQ) can be extremely small, e.g., reading/writing a pointer value (64 bits or 8\unit{B}).
As we discuss in the next section, presence of such small messages makes it extremely challenging for existing Ethernet fabrics to provide low latency with high bandwidth utilization for memory traffic.

%\begin{figure}
%    \centering
%    \includegraphics[width=0.475\textwidth]{figures/arch1.pdf}
%    \caption{Existing end-to-end network stack for memory disaggregation over Ethernet.}
%    \label{fig:arch1}
%\end{figure}

\subsection{Limitations of Existing Ethernet Fabrics}
\label{sec:background:sota}
%Ethernet remains the most dominant communication protocol inside datacenters. 
%Existing Ethernet-based network fabrics, such as TCP/IP and RoCE (RDMA over Converged Ethernet), are built on top of Ethernet Media Access Control (MAC) layer, as illustrated in \autoref{fig:edm_stack}.
%Unfortunately, existing datacenter network stacks are not optimized to support memory flows with low latency and high bandwidth utilization due to the limitations discussed below.
In this section, we discuss the limitations of existing Ethernet fabrics with regard to memory disaggregation. 

\begin{figure}[t!]
    \centering
    \includegraphics[width=0.48\textwidth]{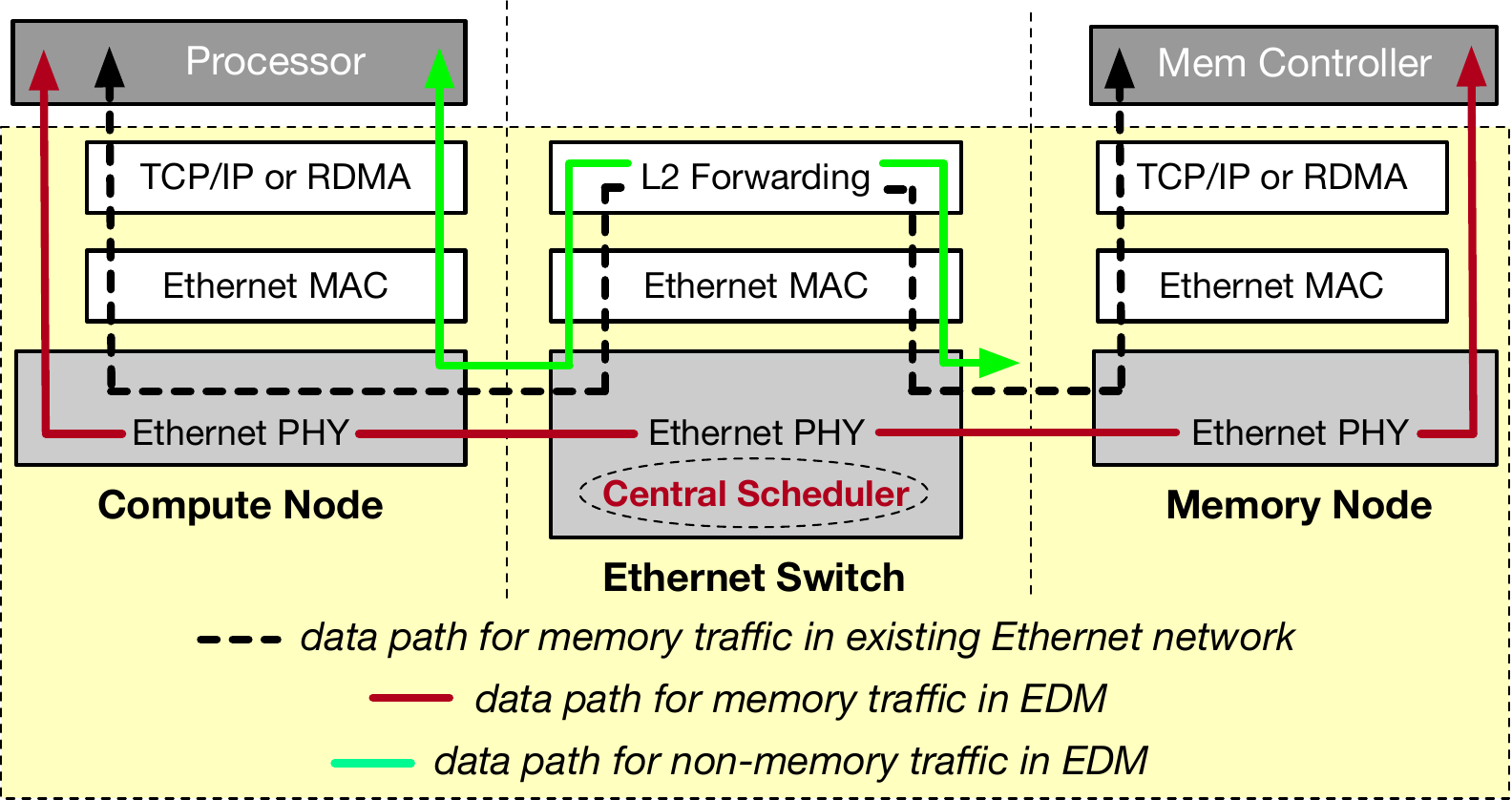}
    \caption{Data path for memory traffic in \sys vs. in existing Ethernet fabrics for memory disaggregation.}
    \label{fig:edm_stack}
    \vspace{-0.25cm}
\end{figure}

First, we consider an \textit{unloaded} network (i.e., no network queuing and congestion). Here, the limitations stem primarily from the communication semantics of Ethernet's MAC layer, on top of which all existing Ethernet-based network protocol stacks for remote memory access, such as TCP/IP and RDMA (RoCEv2~\cite{rocev2}), are implemented (\autoref{fig:edm_stack}).
\begin{itemize}[leftmargin=*]
\setlength\itemsep{0.5em}
    \item \textbf{Limitation 1:} \textit{Minimum frame size overhead.}\\
    Ethernet MAC imposes a minimum frame size of 64\unit{B}. This may result in extremely poor bandwidth utilization for small memory messages, that may be much smaller than 64\unit{B} (\autoref{sec:background:mem_flow}). For example, an 88\% bandwidth wastage while sending 8\unit{B} RREQ messages using minimum-sized Ethernet frames. Batching multiple memory messages destined to the same destination into a single frame may improve utilization, but at the cost of higher latency due to unpredictable wait time to form a batch.
    
    %To improve utilization, one may batch small memory flows into a single frame, but this may result in higher latency for each flow due to unpredictable wait time for batching. Further, batching would only be possible under the special scenario where a source has multiple active memory flows destined to the same destination.
    
    \item \textbf{Limitation 2:} \textit{Inter-frame gap (IFG) overhead.}\\
    IEEE 802.3ae standard for Ethernet enforces a minimum gap of 96 bits between two consecutive MAC layer frames. This results in significant bandwidth overhead for small frames---16\% overhead for 64\unit{B} frames. For traditional datacenter traffic, this overhead is masked by sending larger frames, but for small memory messages, larger frames would demand higher degree of batching to maintain high bandwidth utilization, resulting in higher batching latency.
    
    \item \textbf{Limitation 3:} \textit{No intra-frame preemption.}\\
   % In a practical setting, memory flows will co-exist with traditional datacenter traffic, which may interfere with the memory flows both at the host and the switch. The standard technique used in datacenters to minimize such interference is through \textit{priority classes}, where a latency-sensitive flow (in this case, a memory flow) is given the highest priority for transmission, and hence can preempt the transmission of lower priority non-memory traffic. However, \textit{one cannot preempt the transmission of a single frame at the MAC layer}. While this might be benign for traditional datacenter traffic, for memory flows, this may lead to a high overhead. For example, failure to preempt an 1500\unit{B} MTU-sized packet would add a latency of 120\unit{ns} over a 100\unit{Gbps} link (respectively, 720\unit{ns} for a jumbo Ethernet frame of size 9\unit{KB}). To make matters worse, this interference may happen at each hop along the path of a memory flow.
   In a datacenter setting, memory traffic has to co-exist with non-memory traffic, such as IP and storage. Unfortunately, at the MAC layer, one cannot preempt an Ethernet frame midway through its transmission. %This may lead to high latency overhead for memory traffic. E.g., 
   Failure to preempt a 1500\unit{B} non-memory MTU frame would increase the latency for a remote memory message by 120\unit{ns} over a 100\unit{Gbps} link (resp. by 720\unit{ns} for a 9\unit{KB} jumbo frame). To make matters worse, this interference may happen at each hop, further adding to the remote memory access latency.
    
   % consider a memory flow (e.g., RREQ) arriving while an MTU-sized frame is being transmitted over a 100\unit{Gbps} link. In this case, RREQ will have to wait 120\unit{ns} before it can be transmitted. %This already increases the remote memory access latency by over $2\times$ (over $3\times$ if this happens both at the source and the switch along the path) compared to local memory access latency.
   % To make matters worse, Ethernet also allows jumbo frames that can be 9000\unit{B} in size. Failure to preempt a jumbo frame over a 100\unit{Gbps} link would increase the latency by 720 \unit{ns}. Also note that in practice this interference could happen at any link that is being shared by the memory and non-memory traffic (at the sender, receiver, or switch), and even worse, the same memory frame may experience multiple such delays at different hops along its path, thus bloating the overall remote memory access latency even further.
    %remote memory access latency by over $8\times$ (over $15\times$ if this happens both at the source and the switch) compared to local memory access latency.
    
    \item \textbf{Limitation 4:} \textit{Layer 2 switching overhead.}\\
    %Each Ethernet frame carrying the bytes from a memory flow will be processed and forwarded by a Top-of-Rack layer 2 switch. Inside the switch, the frame goes through several modules~\cite{rmt}, including a parser, one or more match-action stages for table look ups, and a packet manager. Each module adds latency to the pipeline, and the overall latency can be several 100s of ns for state-of-the-art switches (\autoref{table:latency}).
    An Ethernet switch performs layer 2 packet processing to forward frames to their respective destinations. For this, the frame has to go through a switch forwarding pipeline~\cite{rmt} comprising a parser, one or more stages of match-action tables, and a packet buffer manager and scheduler. The total latency of such a pipeline can be several 100s of nanoseconds for modern switches (\autoref{table:latency}).
\end{itemize}

%\smallskip
\noindent
Next, we consider a \textit{loaded} network where frames may get queued and even dropped due to congestion at the switch.\smallskip

\begin{itemize}[leftmargin=*]
\setlength\itemsep{0.5em}
    \item \textbf{Limitation 5:} \textit{Transport layer overhead.}\\
    %Ethernet is a "best-effort" data delivery protocol. As a result, one needs to implement complex reliability, congestion and flow control protocols in the transport layer on top of Ethernet to handle in-network queuing and frame losses. These protocols inevitably add latency to a frame's data path through operations such as adding/extracting the congestion signals, sequence and acknowledgement numbers, receiver's window size, etc. to/from packet headers, that requires header parsing, header encapsulation/decapsulation, and per-flow state updates and look ups. \autoref{table:latency} shows that this could lead to high latency overheads for memory flows.
    %as they carry out operwhich can be quite significant in the context of memory flows (\autoref{table:latency}). %For example, RDMA transport adds a latency of 650\unit{ns} to the data path (325\unit{ns} each on TX and RX), thus increasing the remote memory access latency by over 6.5$\times$ compared to local memory access latency. For TCP/IP transport, it is even worse, as it adds a latency of around 2 $\mu$s to the data path (0.9 $\mu$s on TX and 1.1 $\mu$s on RX), increasing the the remote memory access latency by over 20$\times$ compared to local memory access latency.
    Ethernet relies on a higher-layer transport protocol, e.g., TCP, to provide reliability and handle network congestion.
    Transport protocols add significant latency to a frame's data path (\autoref{table:latency}), which can be prohibitively high for remote memory access. RDMA over Ethernet eliminates some of the reliability overheads by relying on a link level priority flow control (PFC)~\cite{pfc} that guarantees no losses at the switches. However, most practical deployments of RDMA still require a congestion control protocol to avoid bandwidth under-utilization caused by PFC~\cite{dcqcn,timely}.
    
    \item \textbf{Limitation 6:} \textit{Queuing delay and drops at the switch.}\\
    Most datacenter congestion control protocols are reactive in nature~\cite{tcp,dctcp,pfabric,hull,hpcc}, i.e., they rely on some form of congestion feedback to handle congestion. Thus they cannot eliminate queuing, especially for many-to-one (incast) traffic pattern. Memory traffic is extremely sensitive to queuing, where even a few microseconds of queuing delay is prohibitively high. 
    To make matters worse, small memory messages may fit within a single frame, and hence their loss (e.g., due to excessive queuing) would not trigger a fast retransmission via 3 Duplicate ACKs~\cite{tcp}. Instead, timeout (typically set to several $\mu$s~\cite{dctcp,pfabric,ndp}) would be the only recourse.
     %Furthermore, in scenarios where a frame is dropped due to excessive queuing, the latency overhead of retransmission can also be extremely high for small memory flows that may fit within a single frame, and hence their loss would not trigger a fast retransmission, e.g., via 3 Duplicate ACKs~\cite{tcp}. 
     %as that requires multiple frames from a flow to be received at the receiver. Instead, timeout would be the only recourse, which is typically set to a very conservative value, often several $\mu$s~\cite{dctcp,pfabric,ndp}, to avoid redundant retransmissions. 
     The lossless nature of PFC may eliminate the overhead of retransmission, but it does not eliminate queuing delay as PFC is only triggered once the queue at a switch fills up beyond a threshold. 
     %Recent proactive congestion control protocols~\cite{phost,ndp,homa,expresspass,cai2022dcpim}, where receivers carefully schedule flows from the senders, proactively avoid queuing in the network. However, due to their de-centralized nature, these protocols are prone to scheduling conflicts where a sender might receive simultaneous scheduling requests from multiple receivers, but could only respond to one, while ignoring or delaying the rest. This could lead to bandwidth under-utilization and high flow completion time.
     Another class of congestion control protocols, namely proactive congestion control protocols (\autoref{sec:design:fabric}), try to proactively avoid queuing in the network. Such protocols can either be centralized or de-centralized. Unfortunately, de-centralized protocols are either prone to bandwidth under-utilization~\cite{phost,ndp,homa,expresspass,cai2022dcpim} or suffer from poor latency scaling~\cite{sirius,rotornet,shoal}, while the centralized protocols~\cite{fastpass,helios} are bottlenecked by the central server's processing and bandwidth capacity.
     
\end{itemize}

\section{Design}
\label{sec:design}
This section describes the design of \sys. \autoref{fig:edm_stack} shows the architecture of \sys and contrasts it with existing Ethernet fabrics for memory disaggregation. \sys overcomes the limitations of existing Ethernet fabrics, as discussed in \autoref{sec:background:sota}, using two key design ideas:\smallskip\\
\noindent
\circled{\textbf{{\footnotesize D1}}}\hspace{0.1cm}\textbf{Remote memory access protocol in the PHY.} \sys implements the network protocol stack for remote memory access in the Physical Coding Sublayer (PCS) of Ethernet's Physical (PHY) layer. This overcomes the limitations 1--3 mentioned in \autoref{sec:background:sota} as follows: \textit{(i)} PCS operates at a much finer data granularity of 66-bit blocks compared to the 64\unit{B} minimum frame size at the MAC layer. This avoids bandwidth wastage for small memory messages; \textit{(ii)} PCS has access to the inter-frame gap (IFG) bits, which one could repurpose to carry memory messages, thus eliminating the bandwidth overhead of IFG; and \textit{(iii)} Operating at 66-bit granularity inside the PCS also enables preemption of MAC layer Ethernet frames, thus reducing the impact of large non-memory frames on the latency of small memory messages.   
%\textit{intra-frame preemption} in PHY, where one may preempt the transmission of a large frame from non-memory traffic in favor of small memory message for low latency. %and \textbf{(iv)} By switching at the PHY layer one also overcomes the latency overheads of layer 2 switch processing and forwarding.
\smallskip\\
\noindent
\circled{\textbf{{\footnotesize D2}}}\hspace{0.1cm}\textbf{Centralized in-network memory traffic scheduler.} 
This overcomes the limitations 4--6 mentioned in \autoref{sec:background:sota}. First, a central scheduler has a global view of all active memory traffic and hence can optimally schedule the traffic such that there is zero queuing at the switch while ensuring high bandwidth utilization (overcoming limitation 6). Furthermore, with no queuing and congestion in the network, the transport layer at the hosts does not need to implement complex reliability, flow and congestion control protocols (overcoming limitation 5).
Finally, a central scheduler avoids contention by creating virtual circuits between sources and destinations. \sys implements the scheduler in the PHY of the switch. Thus, memory traffic is forwarded through the virtual circuits in the PHY, avoiding layer 2 packet processing overhead on the switch (overcoming limitation 4).

\smallskip
In the following sections, we describe the above two design ideas in more detail.

\subsection{In-Network Scheduler}
\label{sec:design:fabric}

\mypar{Context} \sys's scheduler falls within the class of proactive network traffic schedulers that try to \textit{proactively} avoid queuing and congestion in the network. The core idea behind all existing proactive network schedulers is to form an online maximal matching~\cite{edge-matching} between sources and destinations (for high bandwidth utilization) over disjoint network paths (to avoid contention and queuing). %The distinction between different schedulers comes from the choice of architecture (e.g., centralized vs. de-centralized) and scheduler hardware target (e.g., server vs. switch).
The scheduler could either be centralized or de-centralized.

Receiver-driven protocols~\cite{phost,ndp,homa,expresspass,cai2022dcpim} are examples of de-centralized proactive schedulers, where receivers carefully schedule data from the senders to proactively avoid queuing at the network edge. However, due to their de-centralized nature, these protocols are prone to scheduling conflicts where a sender might receive simultaneous scheduling requests from multiple receivers but could only respond to one, while ignoring or delaying the rest. This could lead to bandwidth under-utilization and a long flow completion time (\autoref{sec:eval}). Another class of de-centralized proactive schedulers, used primarily in reconfigurable circuit-switched networks~\cite{sirius,shoal,rotornet,opera}, use a static pre-defined set of repeating matchings for scheduling. Unfortunately, such designs have extremely poor latency scaling w.r.t. number of nodes~\cite{shoal,orn_stoc}. %Further, the use of a static schedule fundamentally dictates that any improvement in the latency scaling (e.g., by using a different set of pre-defined matchings) would come at the cost of network throughput~\cite{orn_stoc}.

Centralized schedulers use the knowledge of the global network traffic demand to overcome the limitations of de-centralized solutions, at the cost of scalability. These schedulers typically target rack- or cluster-scale deployments, similar to \sys. Existing centralized schedulers are implemented in software on a remote server~\cite{fastpass,helios} or in hardware on a layer 2 programmable switch~\cite{pl2}. Unfortunately, server-based software solutions take tens of $\mu$s to make scheduling decisions~\cite{fastpass}, thus incurring high latency overhead. Furthermore, the remote server architecture does not scale to clusters with high aggregate bandwidth or to workloads with a large number of small messages (such as memory traffic), as the server bandwidth is easily overwhelmed by the control messages carrying the traffic demand (\autoref{sec:eval}). On the other hand, programmable switch-based solutions could only implement an approximate matching due to the computational limitations of a programmable switch~\cite{pl2}, not to mention the added latency due to layer 2 packet processing (\autoref{sec:background:sota}).

\mypar{Our contribution} \sys presents a new design point in the space of proactive centralized network schedulers, by proposing a custom hardware pipeline for the scheduler in the PHY of an Ethernet switch. As discussed in the following, this design overcomes both the latency and bandwidth limitations of server-based centralized schedulers, and the computational constraints and layer 2 packet processing overheads of programmable switch-based schedulers.

%\sys implements a central scheduler for memory messages in the Top-of-Rack switch that acquires a global knowledge of all the active memory messages in the network, and optimally schedules them over the network. %The overall performance goal of the scheduler is to minimize the completion time of each memory message while maximizing the bandwidth utilization.

\subsubsection{Scheduler design}
\label{sec:design:fabric:scheduler}
 \sys's scheduler has two components: \textit{notification} (used to acquire the global memory traffic demand) and \textit{grant} (used to schedule memory traffic using an online maximal matching algorithm). At a high level, senders notify the switch of any memory messages they want to send over the network and then wait for the switch to grant them the permission to send. The switch, on the other hand, stores the notifications for all the memory messages from all senders in a single logical queue, called a \textbf{demand notification queue}, and runs an online maximal matching algorithm to schedule memory messages from the notification queue in a way that ensures both high bandwidth utilization and zero queuing. Furthermore, to achieve a near-optimal message completion time under bandwidth contention, \sys extends the maximal matching algorithm with priority-based scheduling, where it assigns a priority value to each memory message and resolves matching conflicts in favor of the message with the highest priority. \sys chooses the optimal priority assignment scheme based on the nature of the memory traffic workload. 
\begin{itemize}[leftmargin=*]
\setlength\itemsep{0.5em}
    \item \textbf{Notification.} For WREQ messages, senders send an explicit demand notification to the switch containing the message's destination and message size, and then wait for a grant before sending the message.
    For RREQ/RMWREQ--RRES message pairs, the RREQ/RMWREQ message serves as an implicit demand notification for the corresponding RRES message. Senders send the RREQ/RMWREQ message to the switch without sending any prior notification. On receiving an RREQ/RMWREQ message from source $s$ destined to $d$, the switch adds the message to the demand notification queue and uses the number of bytes to be read as specified in the RREQ as the demand for RRES message from $d$ to $s$. For RMWREQ, the RRES size is inferred based on the opcode.
    Note that the advanced knowledge of message or flow size is challenging for traditional datacenter traffic, but it is readily available for memory traffic, as a memory access request message must include the number of bytes to be read or written, since it is required by the memory controller interface, such as DDR4.

    \item \textbf{Grant.} The grant mechanism implements a priority-based online maximal matching. A grant from the switch allows the sender to send a maximum of $c$ bytes (called a \textit{chunk}) from a given message over the network. Given a set of messages in the notification queue, a priority value for each message, and a maximum chunk size of $c$ bytes, the scheduler issues grants as follows:
    \begin{enumerate}[label=(\arabic*),leftmargin=*]
    \setlength\itemsep{0em}
        \item Scheduler maintains the count of remaining bytes for each message. Initially this value is set to the message size as communicated in the notification message.
        \item A message with source port $s$ and destination port $d$ is marked as ineligible for scheduling if either $s$ or $d$ is marked {\small \texttt{busy}}.
        Initially all ports are marked {\small \texttt{not\_busy}}, thus making all messages eligible for scheduling.
        \item Scheduler selects an eligible message $m$ with the highest priority from the notification queue.
        \item Assuming $s$ and $d$ are respectively the source and destination ports for message $m$, scheduler sends a grant message to $s$ asking to send a chunk of $l$ bytes $=$ \textit{min}\hspace{0.01cm}($c$, remaining bytes in $m$) from message $m$. It decrements the count of remaining bytes for message $m$ by $l$ bytes. If remaining bytes becomes 0, it removes message $m$ from the notification queue. Note that if $m$ corresponds to an RRES message, then the first grant message for $m$ will be the corresponding RREQ message (that was buffered in the notification queue). 
        \item Scheduler marks $s$ and $d$ as {\small \texttt{busy}}, thus making messages with source $s$ or destination $d$ ineligible for scheduling.
        \item Scheduler keeps repeating steps (3)---(5) forever.
        \item Asynchronously, after receiving the $l$ bytes from a message, scheduler marks the corresponding source and destination as {\small \texttt{not\_busy}}, thus making some of the previously ineligible messages eligible for scheduling. However, waiting till all the $l$ bytes have been received before marking the source and destination as {\small \texttt{not\_busy}} would result in bandwidth under-utilization. This is due to non-zero propagation delay for grant messages. Ideally, we want to send the grant for the next chunk at time $t$ such that the first bit of the next chunk arrives at the switch right after the last bit of previous chunk has been received. Assuming one-hop propagation delay of $P$, link bandwidth of $B$, and transmission delay of grant message as $T$, the first bit of the next chunk will arrive at the switch at time $t+T+2P$. Whereas, the last bit of the previous chunk will arrive at time $t'+T+2P+l/B$, where $t'$ is the time the grant for the previous chunk was issued. Hence, $t$ must be $t'+l/B$, i.e., we must mark the source and destination of message $m$ as {\small \texttt{not\_busy}} after $l/B$ time units have elapsed since the grant for $m$ was issued.
    \end{enumerate}
    \end{itemize}
    \noindent
    \sys's scheduler design leads to following key properties:
    \begin{enumerate}[leftmargin=*]
    \setlength\itemsep{0.25em}
        
        \item \textbf{Maximal matching with zero queuing.} The grant algorithm implements the well-known greedy algorithm for online maximal bipartite matching~\cite{edge-matching}. Maximal matching leads to near-optimal bandwidth utilization in practice~\cite{pim}, as it ensures that a message is not scheduled only if either its source or destination is already busy sending or receiving other messages. Another consequence of matching is it guarantees that there will be at most one source sending to a destination at any given time, thus resulting in zero queuing and no drops at the switch.

        \item \textbf{Zero processing delay during forwarding.} Whenever the scheduler issues a grant for message $m$: $s$$\rightarrow$$d$, it practically creates a virtual PHY circuit between $s$ and $d$. Hence, the data from message $m$ does not need any processing at the switch during forwarding, e.g., no header parsing or table look up as typically needed for layer 2 forwarding.

        \item \textbf{Reduced transport overhead at the host.} With an in-network scheduler that proactively avoids congestion and ensures no data loss due to queuing, one no longer needs a transport stack at the hosts implementing complex end-to-end reliability, flow and congestion control protocols. This further reduces the latency on the data path for memory messages. In \autoref{sec:design:practical}, we discuss how \sys handles losses due to data corruption.
        
        %\item \textbf{Near-optimal message completion time.} Based on prior theoretical results~\cite{opt-scheduling}, given a single global job queue, the optimal scheduling policy to minimize both the average and tail job completion times for light-tailed job size distributions is First-Come-First-Serve (FCFS) and for heavy-tailed job size distributions is Shortest-Remaining-Processing-Time (SRPT). \sys already stores message notification queue as a single logical priority queue. Thus, by simply setting the priority value for each message appropriately in the grant algorithm, \sys can implement both FCFS and SRPT algorithms. For light-tailed message size distributions, \sys will set the priority value for each message as its time of notification, thus implementing FCFS. And for heavy-tailed message size distributions, \sys will set the priority value for each message as remaining bytes in the message (a state that the scheduler already maintains), thus implementing SRPT.

        \item \textbf{Near-optimal message completion time.} The message priority value in \sys can be set appropriately to achieve near-optimal message completion time for different workloads. For light-tailed workloads, \sys sets the priority value of a message as the time of its notification, to implement First Come First Serve (FCFS) scheduling, known to be optimal for light-tailed workloads~\cite{opt-scheduling}. For heavy-tailed workloads, \sys sets the priority value of a message as the remaining bytes in the message (a state that the grant algorithm already maintains), to implement Shortest Remaining Processing Time (SRPT)~\cite{srtf} scheduling, known to be optimal for heavy-tailed workloads~\cite{opt-scheduling}.

        \item \textbf{In-order message delivery.} \sys ensures that the memory messages between a compute--memory node pair are delivered in-order. In general, a non-FCFS scheduling policy such as SRPT may result in out-of-order delivery, but \sys guards against this by applying the SRPT policy only across messages from different compute--memory node pairs. As a consequence, \sys does not provide any ordering guarantees for messages originating from different compute nodes. \sys's ordering semantics are consistent with existing Ethernet-based memory disaggregation systems, such as RDMA-based systems, as well as with the assumptions made in modern multi-processing and distributed systems that only assume that the memory requests from a given process (or node) will be scheduled in-order but do not make any scheduling (or ordering) assumptions for memory requests across processes (or nodes). Instead, they rely on the applications to use explicit synchronization primitives, such as locks and mutexes, to ensure consistency~\cite{dragojevic2014farm}. \sys can support such synchronization primitives using the RMWREQ messages, as described in \autoref{sec:design:stack:host}.
    \end{enumerate}

     %\mypar{Handling host interconnect bandwidth bottleneck} The above design implicitly assumes that the network link bandwidth is the bottleneck and not the bandwidth of the host interconnects (e.g., bandwidth of local memory interconnect). While this is true for existing systems, in the future it may be possible that the Ethernet link bandwidth exceeds the host interconnect bandwidth. One could easily extend \sys's design to accommodate this, by allowing each host to periodically send its local queuing information to the switch, and the switch would mark the port connected to a host as \texttt{busy} in case the queue occupancy exceeds a threshold, thus ensuring that the requests to that host are not scheduled until the queue at the host has drained.

\subsubsection{Scheduler hardware pipeline}
\label{sec:design:fabric:hardware}
The primary data structure that the scheduler maintains is the demand notification queue. The notification queue must be a priority queue to implement priority-based scheduling policies such as SRPT.  Logically, the scheduler maintains a single global notification queue. But in practice, this would bottleneck the rate of insertions, as one could only add one notification per clock cycle to a single queue, but the switch may receive anywhere between 0 and $N$ notifications in a given clock cycle, where $N$ is the number of switch ports. A single queue also makes it challenging to perform parallel operations on the queue elements. Hence, \sys maintains $N$ notification queues, one per destination switch port. The notification queue for the destination port $d$ stores notifications for messages whose destination host is connected to the port $d$. 
\sys bounds the size of per port notification queues to $X*N$. Here, $X$ is the maximum number of active notifications allowed per source--destination pair. This is achieved by senders rate limiting their active notifications to at most $X$ per destination. As an optimization, senders may also batch several small active messages between a source--destination pair into one "mega" message and send 1 notification for that mega message. In our evaluations, we empirically find that the value of $X$=3 works best (\autoref{sec:eval:simulations}).
Next, \sys uses recent hardware data structures for ordered lists~\cite{pifo,pieo,mp5,thanos} to implement its notification queues. These data structures could perform priority queue operations in a constant number of clock cycles. In particular, the latency of inserts and deletes is 2 clock cycles, with both operations fully pipelined, i.e., one may issue a new operation every clock cycle. The latency to get the highest priority element in the queue is 1 clock cycle.

Next, the scheduler must implement an online maximal matching over the set of messages in the notification queue. A naive implementation of the matching algorithm, as described in Steps (3)---(5) of the grant algorithm (\autoref{sec:design:fabric:scheduler}), would match one source--destination pair each clock cycle. This would take $\sim$$N$ clock cycles to form a maximal matching. Instead, \sys borrows inspiration from the classic Parallel Iterative Matching (PIM)~\cite{pim} algorithm, which performs the matching for each destination switch port in parallel.

PIM is an iterative algorithm that forms a matching in each iteration and on average takes $\sim$$log(N)$ iterations to form a maximal matching. \sys extends PIM's design to incorporate message priorities to implement scheduling policies such as SRPT. \sys implements each iteration of priority-based PIM in exactly 3 clock cycles. In the first clock cycle of each matching iteration, each destination port $d$ in parallel chooses the highest priority eligible message $m$: $s$$\rightarrow$$d$ from its respective notification queue, i.e., the highest priority message with both $s$ and $d$ marked {\small \texttt{not\_busy}}, and issues a request to form the corresponding matching between $s$ and $d$. This takes 1 clock cycle using the fast hardware priority queue data structure for the notification queue. In the second clock cycle, each source port $s$, in parallel, checks if multiple destination ports issued a matching request to $s$, and if so, it generates a grant for the message $m_j$: $s$$\rightarrow$$d_j$ with the highest priority, while discarding the remaining matching requests. In the third clock cycle, $s$ and $d_j$ are marked as {\small \texttt{busy}} and we move on to the next iteration of PIM. %Thus, each iteration only takes 3 clock cycles.

One of the challenges in the above design is to find the highest priority request from the set $M$ = $\{$$m_i$: $s$$\rightarrow$$d_i$$\}$ of all matching requests for a source port $s$, during the second cycle of each PIM iteration. Na\"{\i}vely, this would take $log(n)$ clock cycles, where $n$ is the number of requests in the set $M$. However, \sys does this in exactly 1 clock cycle by intelligently trading off hardware resources for time. In particular, \sys maintains an array of size $N$ per source port $s$. The array stores the destination port numbers $d_i$ sorted by the highest priority value in each $d_i$'s notification queue, and a boolean value for each array index, initialized to 0. The array is implemented using the same ordered list data structure as the notification queue, and just like the notification queue, it is updated on the arrival of each demand notification or whenever the priority of a message changes (e.g., when the remaining message size changes in SRPT). Given this, during the second cycle of a matching iteration, for each source port $s$ in parallel, each destination port $d_i$ with the matching request for message $m_i$: $s$$\rightarrow$$d_i$ sets the boolean value in the array index storing $d_i$ to 1 in parallel, while a priority encoder~\cite{pri-encoder} synchronously returns the most significant index in the array set to 1, which, in turn, corresponds to the destination port with the highest priority matching request.

%\subsubsection{Analysis}
%\label{sec:design:fabric:overhead}
%\smallskip
%\vspace*{0.1cm}
\subsubsection{Scheduling latency and throughput} 
As discussed in \autoref{sec:design:fabric:hardware}, \sys takes $log(N)$ iterations on average to form a maximal matching. Given that \sys implements each iteration in 3 clock cycles, it takes $3$ $*$ $log(N)$ clock cycles on average to form a maximal matching. Given a clock rate of $R$\unit{GHz} for the scheduler hardware pipeline (\autoref{sec:design:fabric:hardware}), the latency to form a maximal matching is $T=3*log(N)/R$\unit{ns}. Hence, to achieve line rate scheduling, \sys needs to set the minimum chunk size to $T$\unit{ns} transmission time, to ensure that the link remains busy while forming the next maximal matching. Our scheduler design is estimated to run at 3\unit{GHz} on an ASIC (\autoref{sec:eval:prototype}), thus needing only 9\unit{ns} on average to form a maximal matching for a 512-port switch. Hence, to achieve line rate scheduling for 512$\times$100\unit{Gbps} switch, \sys would set the minimum chunk size to 128\unit{B}.

\subsubsection{Network bandwidth and latency overhead}
\label{sec:design:fabric:overhead}
%Both notification and grant messages are small, thus limiting their bandwidth overhead. 
Assuming $m$ is the memory message for which notification/grant is issued, both notification and grant message will contain the destination of $m$ (9 bits for a cluster of size 512), the message id of $m$ used to distinguish messages between the same source--destination pair (8 bits), and size of $m$ for notification or chunk size for grant (16 bits). In addition, a notification is sent once for every WREQ message, while a grant is sent once for every chunk. For a chunk size and WREQ message size equal to the burst size for DDR4 (64\unit{B}), the bandwidth overhead for both grant and notification is limited to 6\%. This is further reduced for larger chunk/message sizes. %However, increasing the chunk size to reduce the bandwidth overhead for grant messages may come at the cost of performance, as smaller chunk sizes enable a more fine granular scheduling and message multiplexing. Empirically, a chunk size of $\sim$1/2 BDP leads to the optimal trade-off in \sys, and we set the chunk size to this value in our evaluations (\autoref{sec:eval:simulations}).

The latency overhead for WREQ messages is $\sim$RTT/2, i.e., the time it takes to send a notification message to the switch and receive a grant message back, before sending the actual WREQ message. In a rack or cluster setting where RTTs are extremely small, this overhead is only a few nanoseconds, which is a small price to pay to avoid queuing delays at the switch. %We include this latency overhead for remote writes in our evaluation in \autoref{sec:eval}. 
Further, for RREQ, RMWREQ, and RRES messages, the above latency overhead does not apply, as the switch treats the RREQ/RMWREQ messages as implicit demand notifications for the corresponding RRES messages, and the memory node treats the received RREQ/RMWREQ messages from the switch as implicit grants for RRES messages.

\begin{figure*}
    \centering
    \includegraphics[width=.95\textwidth]{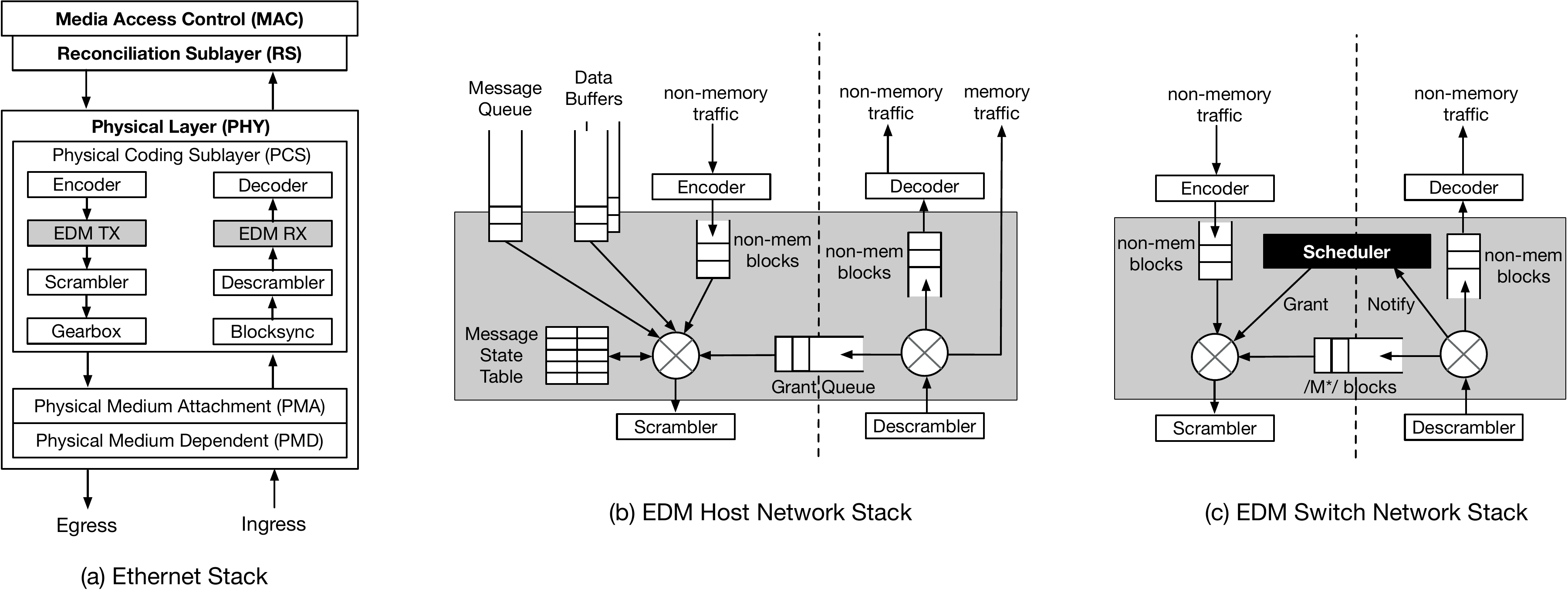}
    \caption{\sys network stack.}
    \label{fig:stack}
\end{figure*}

\subsection{Network Stack}
\label{sec:design:stack}

In this section, we describe the design of \sys's host and switch network stack for remote memory access.

\sys’s network stack  operates in the Ethernet PHY. As shown in \autoref{fig:stack}, Ethernet PHY has three components: Physical Coding Sublayer (PCS), Physical Medium Attachment (PMA), and Physical Medium Dependent (PMD). \sys's logic sits in the PCS layer between the encoder and the scrambler modules on the transmit (TX) path and between the decoder and the descrambler modules on the receive (RX) path. The reason for this architecture is two-fold. First, in the 10/25/40/100+ GbE standard, the encoder module in the PCS layer reformats the input frame on the TX path into a sequence of 66-bit blocks. Thus, the interface between the encoder and the scrambler on the TX path and the descrambler and the decoder on the RX path is 66-bit blocks. Hence, by sitting in between, \sys is able to operate at a fine-grained data transfer unit of 66 bits. Second, the encoder also adds the idle characters (all 0s by default) to form the inter-frame gap (IFG), and the decoder subsequently removes them on the RX path. Thus, by sitting at the output of the encoder (and input of the decoder), \sys is able to utilize the bandwidth otherwise used by idle characters to carry memory messages~\cite{dtp2016,dtp2019,covert,bw_est,seer}. In order to comply with the Ethernet standard, on the RX side, \sys extracts the memory messages and replaces them with idle characters (all 0s) before sending them to the decoder.

%Note that using idle characters to exchange memory traffic does not violate the Ethernet standard for IFG~\cite{dtp2016,dtp2019,covert,bw_est,seer}, as \sys still ensures 96 bits of minimum gap between layer 2 Ethernet frames; the only difference now being that instead of all 0s, they carry memory data, which we subsequently extract and replace with 0s at the RX before sending to higher layers.

A 66-bit PHY block output from the encoder comprises 2 bits of sync header and 64 bits of block payload. An Ethernet frame starts with an /S/ control block, followed by several /D/ data blocks, and ends with a /T/ control block. Ethernet also has a special /E/ control block that is typically used to make up the IFG. Each control block has a sync header value of "01" and an 8-bit block type, followed by 56 bits of payload (which could be frame data). The data block has sync header value of "10" and 64 bits of payload containing frame data. Ethernet enforces at least 9 PHY blocks (/S/, /T/, 7 /D/ blocks) per frame to meet the minimum Ethernet frame size.

In order to distinguish memory messages from normal Ethernet traffic to provide differentiated service in the PHY, \sys introduces a new set of block types collectively called /M*/ blocks. An /MD/ block is identical to the /D/ data block except that it carries memory data. An /MS/ control block marks the start of a memory message, and an /MT/ block marks the end. \sys also introduces an /MST/ control block to support memory messages that may fit within a single PHY block. Thus, unlike an Ethernet frame that must comprise at least 9 PHY blocks, a memory message in \sys can be as small as a single PHY block. \sys also introduces an /N/ and a /G/ control block to send demand notifications and grant messages, respectively (\autoref{sec:design:fabric:overhead}). To distinguish \sys control blocks from each other and from standard Ethernet control blocks, we assign them unique unused block-type values.

%Next, we describe the design of \sys's network stack.

\subsubsection{Host network stack}
\label{sec:design:stack:host}
\autoref{fig:stack} shows the architecture of \sys's host network stack.
\sys's data path runs in parallel with the standard Ethernet data path carrying non-memory traffic. In order to distinguish remote memory messages from non-memory messages (e.g., IP and storage), \sys introduces four new APIs to generate memory messages (RREQ, WREQ, RMWREQ, RRES), as described in \autoref{sec:background:mem_flow}. 

On the TX side, the generated RREQ/RMWREQ/WREQ messages are stored in a \textit{message queue}, along with a pointer to a \textit{data buffer} storing the data to be written (for WREQ).
\sys continuously dequeues messages from the message queue. \sys assigns each dequeued message a \textit{message id} in order to distinguish messages destined for the same destination. For the dequeued RREQ/RMWREQ message, \sys sends the message along with its message id to the switch using /M*/ blocks. It also adds an entry to the \textit{message state table}, indexed by <message destination, message id>, containing the local memory address to which the corresponding RRES must be written. For the dequeued WREQ message, \sys creates an /N/ block (\autoref{sec:design:fabric:overhead}) and sends it to the switch, while storing the corresponding WREQ state (remote memory address, data buffer address) in the message state table.

On the RX side, when receiving a /G/ block, \sys adds it to a \textit{grant queue}, while when receiving RREQ /M*/ blocks (at the memory node), \sys sends the blocks to the memory controller and adds an entry to the grant queue to generate the corresponding RRES message. Once the data corresponding to an RREQ has been read from the memory, a corresponding entry is added to the message state table pointing to the data buffer storing the read data. Asynchronously, the TX side dequeues from the grant queue, looks up the corresponding message entry in the message state table for the remote memory write address and data buffer address, reads a chunk of bytes from the data buffer as requested in the grant, and generates a WREQ message (at the compute node) or an RRES message with the message id of the corresponding RREQ (at the memory node) using /M*/ blocks. %If there is no entry for an RRES message in the message state table, \sys checks the notification queue. If no such entry exists in the notification queue as well, \sys defers the grant assuming that the data have not yet been read from the memory. Else, \sys dequeues the RRES message from the notification queue, stores the relevant state information in the message state table, and sends out the requested chunk of data using the /M*/ blocks.
On receiving the /M*/ data blocks corresponding to an RRES message, \sys writes them to the local memory after reading the corresponding memory address from the message state table. On receiving the /M*/ data blocks corresponding to a WREQ message, \sys sends them directly to the memory controller.\smallskip\\ %Finally, \sys converts each received /M*/ block into an /E/ block before sending to the decoder, thus ensuring higher layers are unaffected by \sys.
\noindent
\textbf{Implementing RMWREQ.}
On receiving an RMWREQ message from a compute node, the NIC at the memory node issues the corresponding read request to the local memory controller, followed by the modify operation on the read value as dictated by the opcode. Finally, the NIC writes the modified value to local memory. All three steps are performed atomically, i.e., they are not preempted by other incoming memory requests. Once the atomic read-modify-write operation is finished, the memory node may respond with an RRES message. \sys uses this architecture to implement an atomic compare-and-swap (CAS) for locks and mutexes.\smallskip\\
\noindent
\textbf{Latency of \sys host processing.} Generating an /N/ or an RREQ /M*/ block takes 2 clock cycles---reading from the message queue (1 cycle) and creating the block while writing to the message state table in parallel (1 cycle). Reading a grant from the grant queue takes 4 clock cycles, as grant queue crosses the RX and TX clock domains. Generating an /M*/ data block for an RRES/WREQ message takes 3 clock cycles---reading from the message state table (1 cycle), reading from the data buffer (1 cycle), creating the block (1 cycle). Processing a received /G/ block takes 2 clock cycles---parsing (1 cycle) and adding to grant queue (1 cycle). Processing a received RREQ /M*/ block takes an additional 1 clock cycle for sending to the memory controller. Processing a received /M*/ block from an RRES/WREQ message takes 3 clock cycles---parsing (1 cycle), extracting the memory address (1 cycle), sending data to the application/memory controller (1 cycle).

\subsubsection{Switch network stack}
\label{sec:design:stack:switch}

\autoref{fig:stack} shows the architecture of \sys's switch network stack.
A grant generated by the scheduler is sent to the corresponding host using a /G/ block. Generating a /G/ block takes 1 clock cycle.
On receiving a block, it takes 1 clock cycle to identify /N/, /G/, /M*/ blocks by checking the block type.
When receiving an /N/ block or an /M*/ block(s) corresponding to an RREQ/RMWREQ message, \sys buffers them in the notification queue (\autoref{sec:design:fabric:scheduler}). Instead, on receiving /M*/ blocks from WREQ or RRES messages on the RX of an ingress port, \sys forwards them to the TX of the corresponding egress port. The mapping of RX and TX for each message is set up by the scheduler during the grant (\autoref{sec:design:fabric:scheduler}), so no processing is done on the received /M*/ blocks for forwarding. The latency of forwarding is 4 clock cycles to account for the movement of data from the RX to the TX clock domain.

\subsubsection{Reducing interference of non-memory traffic}
\label{sec:design:stack:isolation}
In converged Ethernet, memory traffic co-exists with traditional non-memory traffic (e.g., IP and storage). As discussed in \autoref{sec:background:sota}, the inability to preempt the transmission of an Ethernet frame at the MAC layer may result in non-memory traffic interfering with the memory traffic sharing the same link, potentially increasing the latency for memory traffic. To reduce this interference, \sys presents a novel design of intra-frame preemption, enabled by PHY processing.

\mypar{Intra-frame preemption} For each link, on the TX path, \sys maintains a buffer at the output of the encoder module in the PCS layer to store the 66-bit PHY blocks from non-memory traffic. By default, \sys implements fair scheduling to schedule the transmission of memory (/N/ or /M*/) and non-memory blocks (although in principle one could use any scheduling policy, e.g., strictly prioritizing memory blocks over non-memory blocks). The scheduling granularity of 66 bits ensures that a memory block is not blocked for the entire duration of transmission of a non-memory frame. 
%\sys dequeues a block from this buffer for transmission only when there are no /N/ or /M*/ blocks ready for transmission. 
However, a consequence of this design is that the PHY blocks in a non-memory Ethernet frame may arrive in non-consecutive cycles at the receiver (but still in order). Unfortunately, the decoder module in PCS and the Ethernet MAC layer expect the blocks from a frame to arrive in consecutive clock cycles for their correct operation. To enforce this, on the RX side, \sys maintains a buffer at the input of the decoder module to buffer the blocks from a non-memory Ethernet frame until it receives a /T/ block, marking the end of that frame. After that, \sys sends the blocks from the buffer in consecutive clock cycles to the higher layers. The size of this buffer is bounded by the maximum-sized Ethernet frame. The buffering does add a nominal latency on the RX data path for non-memory frames even in the absence of memory messages, equal to the transmission delay of that frame. On the TX side, \sys bounds the non-memory buffer size in an event of preemption by sending a back-pressure all the way up to the application's interface to the MAC layer. The latency of this data path is deterministic and equal to 4 clock cycles, thus bounding the buffer size to 4 PHY blocks.

\subsection{Practical Concerns}
\label{sec:design:practical}
This section discusses several practical concerns with \sys.

\mypar{Interoperability}
\sys modifies the Ethernet PHY in both the NIC and the switch. To access remote memory using \sys, one would need an \sys-capable NIC at both the compute and the memory node, as well as an \sys-capable switch in between. However, it is important to note that \sys does not replace the standard Ethernet pipeline on these devices; it simply creates a parallel pipeline to carry memory traffic. Hence, EDM-capable NICs and switches can still communicate with non-EDM Ethernet devices in the network using the standard Ethernet pipeline.
%On both host, modifications to the Ethernet PHY layer in the NIC are necessary to incorporate \sys functionalities. However, it is important to note that an \sys-capable NIC can coexist seamlessly with a traditional NIC for IP traffic transmission. The \sys NIC does not supplant the standard Ethernet pipeline; rather, it implements a parallel data path dedicated to managing memory traffic. 
%On the other hand, likewise, \sys switch shares the same TX and RX with host, besides a flow scheduler. Hence an \sys switch can also fit the any traditional end points.  

%This approach allows for the integration of \sys capabilities while maintaining backwards compatibility with existing network infrastructure, as stated in \autoref{sec:design:stack:host}.

\vspace*{0.1cm}
\mypar{Host integration} \sys's design is oblivious to the specific interconnect technology used in the hosts to connect the NIC to the processor and the memory controller. One may use any low latency interconnect, such as an AXI4 bus (used in our prototype (\autoref{fig:eval_arch})) or one of the many emerging low latency peripheral interconnects~\cite{nanopu,ccix,cxl,nvlink,opencapi,ucie}.% that promise ultra-low latency. 
%In the host stack, \sys sits within the NIC PHY layer, where the notification queue (implemented by BRAM on FPGA) serves as the interface with the host. Thus, while MEM-NIC (e.g., PCIe, CXL.io) paths remain unchanged, \sys requires explicit labeling of remote memory requests. In this way, we assume a transparent translation module sitting alongside with memory module, which intercepts memory instructions and if they are remote requests, \sys requests will be generated and sent to notification queue. Otherwise, the user space program can make direct RDMA-style calls to generate \sys requests. Once the \sys requests arrive on the notification queue in NIC buffer, they will be processed solely by \sys stack.

\vspace*{0.1cm}
\mypar{Application integration} 
For remote memory access, an application may generate \sys messages in \autoref{sec:background:mem_flow} directly (similar to RDMA), or use the traditional load/store API and rely on a shim layer to convert the load/store instructions into the corresponding \sys messages. In both cases, the application will use virtual memory addresses, and a shim layer will intercept all memory requests and perform the virtual to physical memory address translation before directing a request to either the local memory controller or to \sys's stack for remote memory access, depending upon whether the physical memory address is local or remote, respectively.
The design of such a shim layer is beyond the scope of this paper, but there are prior works~\cite{gu2017efficient,ruan2020aifm} that use a similar shim layer for remote memory that could be adapted to \sys.

%In this approach, the shim layer would be responsible for keeping track of which virtual memory address is local and which is remote.
%On the TX side, \sys uses a \textit{notification queue} to interface with the application or the shim layer (at the compute node) and the memory controller (at the memory node).
%The remote memory requests and responses generated by the applications (or the shim layer) and the memory controller are written into the notification queue, along with a pointer to the \textit{data buffer} storing the data to be written or the data read from the remote memory.

\vspace*{0.1cm}
\mypar{Handling data corruption}
While \sys guarantees no data loss due to congestion, there may be loss due to data corruption over a link. In datacenters, data corruption over a link is generally attributed to external factors such as physical damage, bending, or transceiver contamination due to airborne dirt particles, resulting in data transmission errors~\cite{corruption-2}. These errors are not transient, and the only sustainable solution is to disable the damaged link and switch to a back-up network if available (as discussed below) or wait for an operator to fix the link. %especially within a rack where there are typically no alternative links, is to replace the damaged link~\cite{corruption-2}. 
In the Ethernet PHY, the scrambler module checks for data corruption, and if corruption is observed over a link, \sys disables that link.% until an operator fixes or replaces it. %Given that fixing or replacing a link might take time, recently there have also been proposals~\cite{corruption-1} to mitigate losses due to data corruption over a damaged link using link-local retransmission mechanisms. In principle, one could integrate such solutions with \sys to make \sys more robust to data corruption. We leave this as future work.

\vspace*{0.1cm}
\mypar{Fault tolerance}
Within \sys's architecture, the switch is a single point-of-failure. This is no different than any datacenter rack with a single Top-of-Rack (ToR) switch. To guard against such failure, modern datacenter racks typically have a back-up ToR switch network that takes over in case either the primary switch or a link in the primary network fails. In such racks, all hosts have two network interfaces with two links connecting to the two ToR switches. We could also use a similar fault tolerance technique for \sys. However, unlike traditional ToR switches, \sys's switch also stores state for scheduling memory traffic, which will be lost on switch failure, potentially disrupting the entire system's operation. One could guard against this by using the classic state machine replication~\cite{smr}. In particular, on the sender side, \sys would mirror every outgoing remote memory message on both the NIC's interfaces, so that both the primary and back-up switch observe and compute on the same set of messages to have their state synchronized at all times. On the receive side, \sys would accept the first received copy of a memory message and ignore the duplicate received on the other interface. We note that a general state machine replication relies on a consensus protocol~\cite{paxos} to achieve agreement on operation order at each replica. But this is not needed in \sys as all communications happen over a single hop, thus guaranteeing no message re-ordering.

Besides switch and link failure, a memory node failure may also disrupt system's operation. In particular, it may result in a deadlock, where an application at a compute node might block and wait indefinitely for a remote memory read response. Note that such a deadlock would never occur in normal operation, as \sys guarantees no data loss under no failure. However, to ensure the deadlock doesn't happen even under failure, \sys would set a timer for each remote read request, and if the timer expires, \sys would reply to the application with a NULL (zero size) read response.

\section{Evaluation}
\label{sec:eval}

\begin{table*}[h!]
\centering
{\small
\begin{tabular}{|l||c|c||c|c||c|c||c|c|} 
\hline
\multirow{2}{*}{\textbf{Latency Source}} &
  \multicolumn{2}{c||}{\textbf{TCP/IP in hardware}} &
  \multicolumn{2}{c||}{\textbf{RDMA (RoCEv2)}} &
  \multicolumn{2}{c||}{\textbf{Raw Ethernet}} &
  \multicolumn{2}{c|}{\textbf{\sys}} \\
& Read & Write & Read & Write & Read & Write & Read & Write  \\
\hline
\textbf{At Compute Node} & & & & & & & & \\
\hline
Protocol stack & 2$\times$666.2\unit{ns} & 666.2\unit{ns}  & 2$\times$230.2\unit{ns} & 230.2\unit{ns}  & 0 & 0  & 0 & 0       \\ 
\hline
Ethernet MAC   & 2$\times$7.68\unit{ns} & 7.68\unit{ns} & 2$\times$7.68\unit{ns} & 7.68\unit{ns} & 2$\times$7.68\unit{ns}  & 7.68\unit{ns} & 0 & 0       \\ 
\hline
Ethernet PHY (PCS)        & 2$\times$7.68\unit{ns} & 7.68\unit{ns} & 2$\times$7.68\unit{ns} & 7.68\unit{ns} & 2$\times$7.68\unit{ns}  & 7.68\unit{ns} & 2$\times$5.12  & 3$\times$5.12  \\
    & & & & & & & \color{blue}\textbf{+ 12.8\unit{ns}}  & \color{blue}\textbf{+ 28.16\unit{ns}}    \\ 
\hline
\textbf{At Switch}        & & & & & & & & \\ 
\hline
Layer 2 forwarding        & 2$\times$400\unit{ns} & 400\unit{ns}  & 2$\times$400\unit{ns} & 400\unit{ns}  & 2$\times$400\unit{ns}   & 400\unit{ns}  & 0    & 0       \\ 
\hline
Ethernet MAC   & 4$\times$7.68\unit{ns} & 2$\times$7.68\unit{ns} & 4$\times$7.68\unit{ns} & 2$\times$7.68\unit{ns} & 4$\times$7.68\unit{ns}  & 2$\times$7.68\unit{ns} & 0    & 0       \\ 
\hline
Ethernet PHY (PCS)        & 4$\times$7.68\unit{ns} & 2$\times$7.68\unit{ns} & 4$\times$7.68\unit{ns} & 2$\times$7.68\unit{ns} & 4$\times$7.68\unit{ns}  & 2$\times$7.68\unit{ns} & 4$\times$5.12  & 4$\times$5.12  \\
    & & & & & & & \color{blue}\textbf{+ 28.16\unit{ns}} & \color{blue}\textbf{+ 28.16\unit{ns}}    \\ 
\hline
\textbf{At Memory Node}   & & & & & & & & \\ 
\hline
Protocol stack & 2$\times$666.2\unit{ns} & 666.2\unit{ns}  & 2$\times$230.2\unit{ns}  & 230.2\unit{ns} & 0   & 0  & 0    & 0       \\ 
\hline
Ethernet MAC   & 2$\times$7.68\unit{ns} & 7.68\unit{ns} & 2$\times$7.68\unit{ns} & 7.68\unit{ns} & 2$\times$7.68\unit{ns}  & 7.68\unit{ns} & 0    & 0       \\ 
\hline
Ethernet PHY (PCS)        & 2$\times$7.68\unit{ns} & 7.68\unit{ns} & 2$\times$7.68\unit{ns} & 7.68\unit{ns} & 2$\times$7.68\unit{ns}  & 7.68\unit{ns} & 2$\times$5.12  & 5.12  \\
    & & & & & & & \color{blue}\textbf{+ 25.6\unit{ns}}  & \color{blue}\textbf{+ 7.68\unit{ns}}     \\ 
\hline
\hline
\textbf{Network Stack Latency}       & \textbf{3.59~$\mu$s}   & \textbf{1.79~$\mu$s} & \textbf{1.84~$\mu$s}   & \textbf{0.92~$\mu$s} & \textbf{0.92~$\mu$s}   & \textbf{461.44\unit{ns}}  & \textbf{107.52\unit{ns}}  & \textbf{104.96\unit{ns}}      \\ 
\hline
\hline
Ethernet PHY (PMA+PMD) & 8$\times$19 ns & 4$\times$19 ns & 8$\times$19 ns & 4$\times$19 ns & 8$\times$19 ns & 4$\times$19 ns & 8$\times$19 ns & 8$\times$19 ns    \\ 
+ Transceiver delay & & & & & & & & \\
\hline
%\textbf{Transmission Delay}  & 4$\times$20.5\unit{ns}  & 2$\times$20.5\unit{ns}        & 4$\times$20.5\unit{ns}  & 2$\times$20.5\unit{ns}        & 4$\times$20.5\unit{ns} & 2$\times$20.5\unit{ns}        & 2.56+20.5\unit{ns}        & 5.12+20.5\unit{ns} \\ 
%\hline
Propagation delay & 4$\times$10\unit{ns}  & 2$\times$10\unit{ns} & 4$\times$10\unit{ns}  & 2$\times$10\unit{ns} & 4$\times$10\unit{ns}    & 2$\times$10\unit{ns} & 4$\times$10\unit{ns}     & 4$\times$10\unit{ns}        \\ 
\hline
\hline
\textbf{Total Fabric Latency}    & \textbf{3.79~$\mu$s}   & \textbf{1.89~$\mu$s} & \textbf{2.03~$\mu$s}   & \textbf{1.02~$\mu$s}  & \textbf{1.11~$\mu$s}     & \textbf{557.44 ns}    & \textbf{299.52 ns} & \textbf{296.96 ns} \\
\hline
\end{tabular}
\caption{Ethernet fabric latency for remote read and write for four different stacks --- TCP/IP, RoCEv2, Raw Ethernet, and \sys. The one-hop propagation delay is 10\unit{ns}. The latency of TCP/IP and RoCEv2 protocol stacks only include data path latency, i.e., packet header encapsulation/decapsulation and parsing. It does not include latency of control operations, like connection and queue pair set up. The latency for layer 2 forwarding is for a switch programmed to only perform layer 2 forwarding using a single exact match forwarding table [Breakdown -- Parsing (87 ns), Match-Action and table look up (202 ns), Packet Manager (93 ns), Crossbar (18 ns)]. The latency for read is typically double that of write to account for the latency of both RREQ and RRES vs. only WREQ. The overhead of \sys's mechanisms is highlighted in {\color{blue} \textbf{blue}} (fine-grained breakdown in \autoref{fig:breakdown}).}
\label{table:latency}
\vspace{-0.25cm}
}
\end{table*}
We evaluated the performance of \sys using an FPGA-based hardware testbed and large-scale network simulations.
\subsection{Hardware Prototype}
\label{sec:eval:prototype}
We implemented \sys's traffic scheduler (\autoref{sec:design:fabric}) and the host and switch stacks (\autoref{sec:design:stack}) in Verilog by modifying an open-source 25 Gb Ethernet PHY~\cite{corundum}. We synthesized \sys on AMD/Xilinx Alveo U200 FPGA~\cite{alveo} with two network interface ports and 64\unit{GB} off-chip DDR4 DIMMs operating at a total of 77\unit{GB/s}. Further, we envision our switch network stack and the scheduler to eventually be implemented on a multi-port ASIC switching chip. For that, we synthesized a 512-port \sys switch on Synopsys ASIC compiler~\cite{synopsys}, in which \sys's logic takes up only 10 $mm^2$ (1.5--3\% of total area of a modern switching chip~\cite{drmt}), and runs at 3 GHz. The SRAM usage for the notification queue is $K*N^2$ bytes, where $K$ denotes the length of each notification (\autoref{sec:design:fabric:hardware}) and $N$ is the number of ports. For 512 ports, this translates to around 1 MB (1--2\% of total SRAM on a modern switch~\cite{silkroad}).

\mypar{Supporting 100+ GbE} We prototyped \sys in 25 GbE due to the lack of open-source higher-speed Ethernet PHY. However, since Ethernet scales to higher speeds by bundling lower speed PHY lanes~\cite{eth-spec}, e.g., 100GBASE-KR4 is based on 4$\times$25 GbE, our implementation can port to higher speed Ethernet with minimal changes. 

\begin{figure}[t!]
    \centering
    \includegraphics[width=0.48\textwidth]{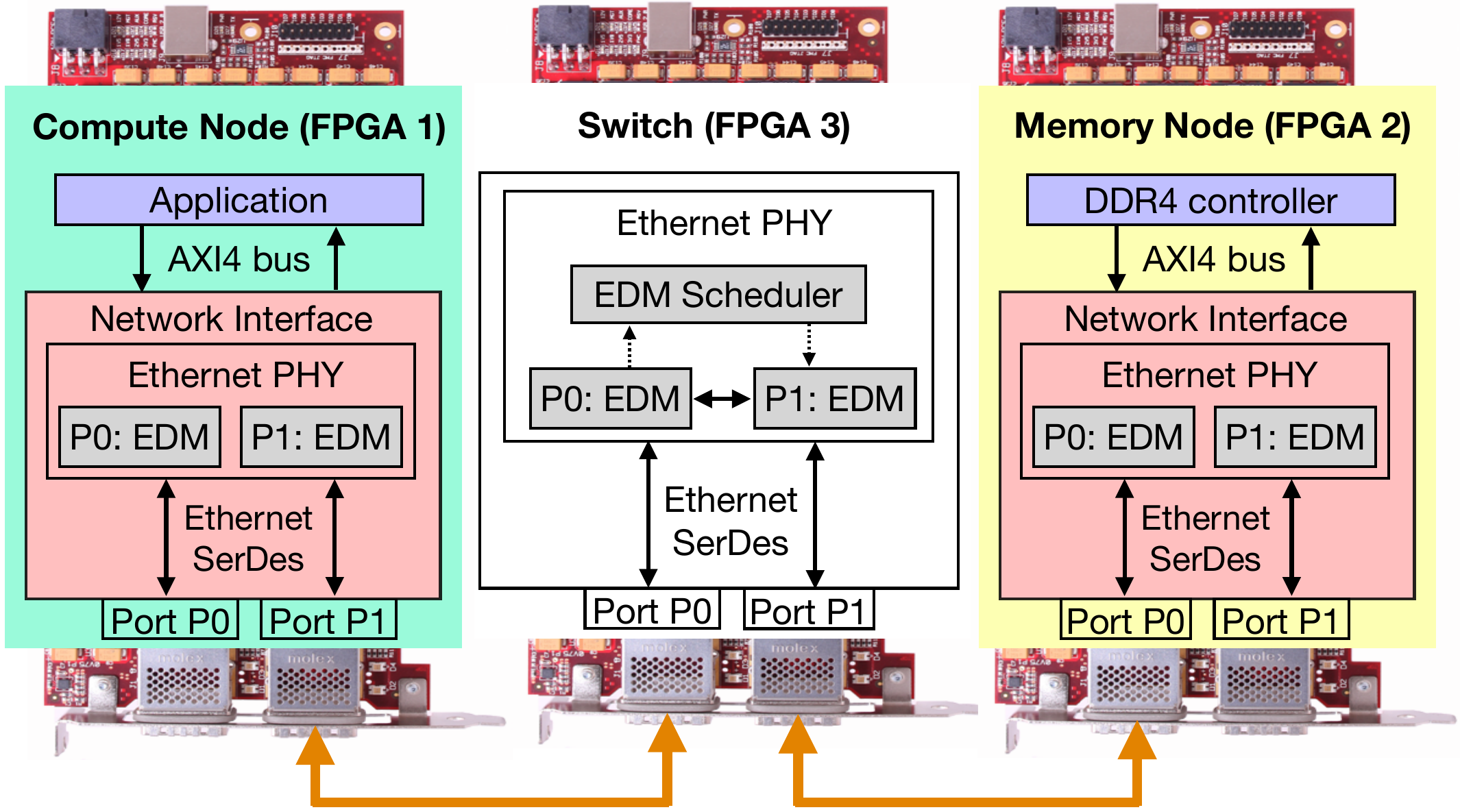}
    \caption{Testbed setup.}
    \label{fig:eval_arch}
    \vspace{-0.25cm}
\end{figure}

\subsection{Testbed Experiments}
\label{sec:eval:testbed:latency}

\begin{figure}[b!]
    \centering
    \includegraphics[width=\columnwidth]{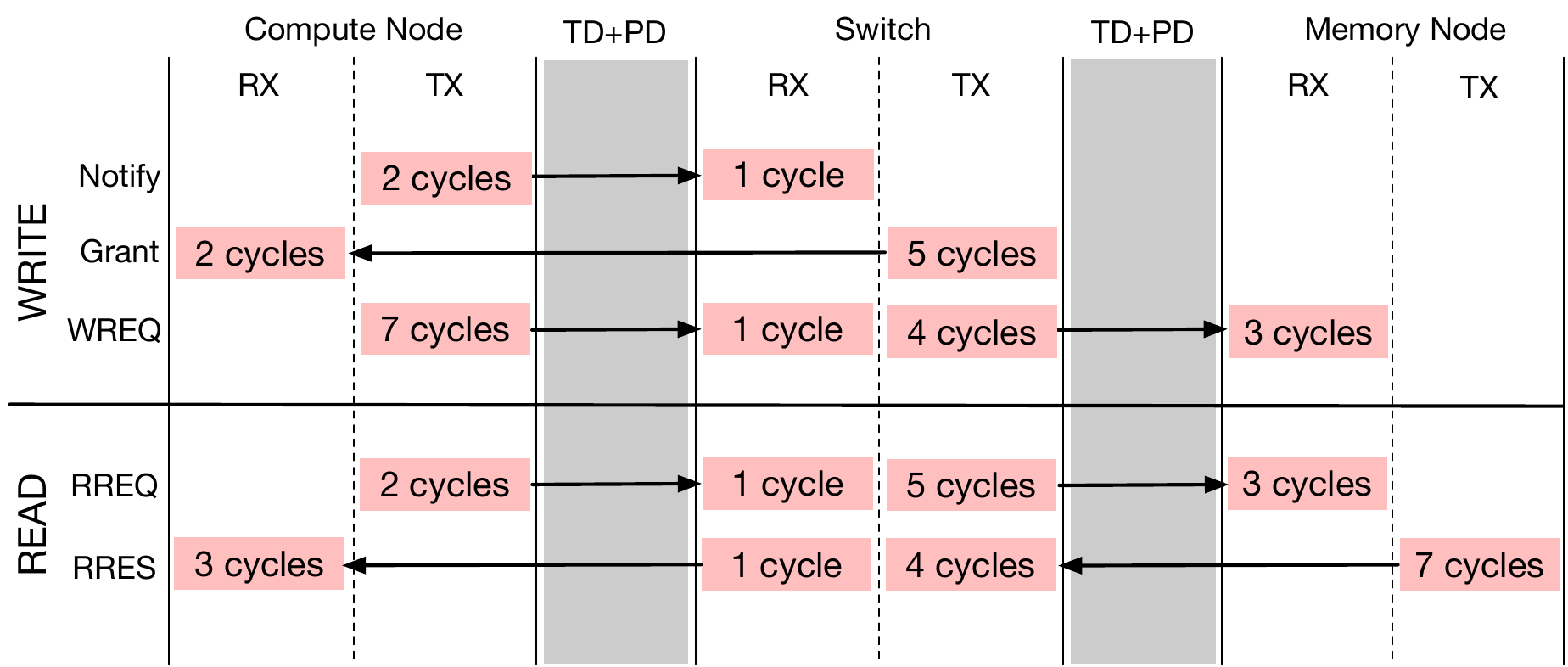}
    \caption{Breakdown of latency for \sys's network fabric for 64\unit{B} read and write. TD+PD = transmission+propagation delay. A clock cycle is 2.56\unit{ns}. For details on the cycle numbers refer to \autoref{sec:design:stack:host} and \autoref{sec:design:stack:switch}.}
    \label{fig:breakdown}
\end{figure}

%In this section, we conduct several microbenchmarks on our FPGA testbed.\smallskip\\
%\noindent
\textbf{Set up.} We connected two \sys FPGAs to a 2-port \sys FPGA switch. We use one FPGA to emulate the memory node, and the other to emulate the compute node (\autoref{fig:eval_arch}). \smallskip\\ 
\textbf{Baselines.} We connected two FPGAs using a layer 2 Tofino switch~\cite{tofino}. We ported three different network stacks to the FPGAs as baselines---\textit{(i)} raw Ethernet (standard Ethernet MAC + PHY only); \textit{(ii)} an open-source FPGA implementation of TCP/IP over Ethernet~\cite{sidler2020strom}; and \textit{(iii)} an open-source FPGA implementation of RDMA over Ethernet (RoCEv2)~\cite{sidler2020strom}.\smallskip\\
\textbf{Workload.} Random reads and writes from compute to memory node. RRES and WREQ are 64\unit{B}, while RREQ is 8\unit{B}.
\subsubsection{Remote memory access latency}
\label{sec:eval:prototype:latency}
In \autoref{table:latency} we report latency in our unloaded testbed. \sys only adds 299\unit{ns} and 296\unit{ns} for remote read and write, respectively. This is comparable to a two hop NUMA latency within a server~\cite{novakovic14sonuma} as well as CXL latency with a single switch~\cite{li2023pond}. Furthermore, the read (write) latency is 3.7$\times$ (1.9$\times$), 6.8$\times$ (3.4$\times$), and 12.7$\times$ (6.4$\times$) lower than the raw Ethernet, RoCEv2 and TCP/IP stacks, respectively. This is a combined consequence of \sys's in-network scheduler and in-PHY stack. \sys's scheduler ensures no congestion and packet drops, thus obviating the need for complex transport mechanisms used in TCP and RoCEv2. Further, \sys's scheduler practically forms a virtual circuit between the source and destination in the PHY, thus eliminating the need for layer 2 packet processing at the switch, which includes parsing, table look-up, buffering, scheduling, and store-and-forward. The trade-off is a few extra cycles in the PHY for \sys.
Additionally, \sys's novel intra-frame preemption mechanism (\autoref{sec:design:stack:isolation}) ensures that the non-memory traffic has minimal interference on the memory traffic. Our testbed experiments showed that even under interference from IP traffic, \sys maintained a near-constant $\sim$300\unit{ns} remote memory access latency.

\begin{figure}[b!]
    \centering
   \includegraphics[width=.95\columnwidth]{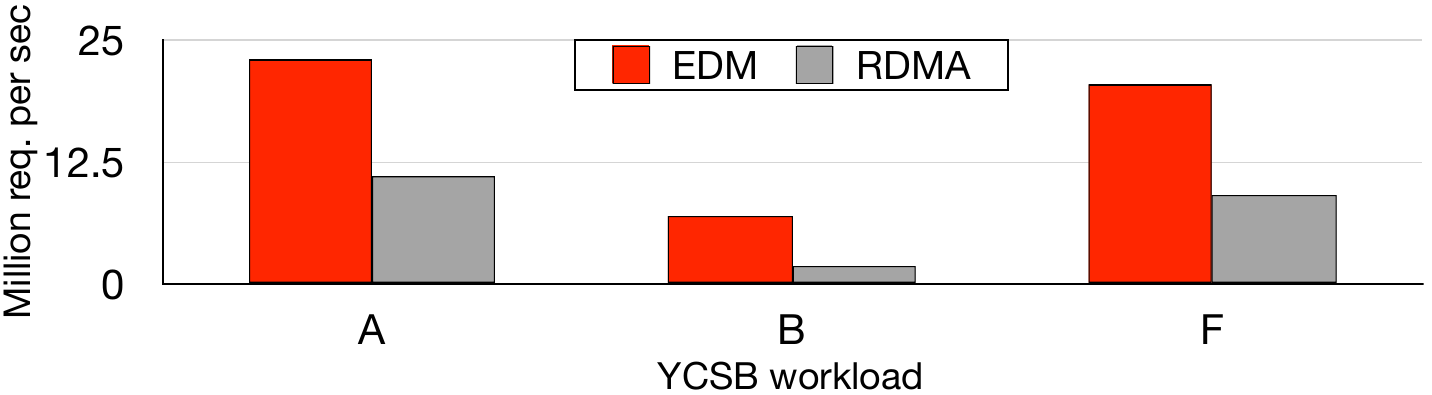}
  \caption{Illustrates a PHY protocol such as \sys can achieve much higher bandwidth utilization (throughput) compared to protocols implemented on top of Ethernet MAC (RDMA).}
    \label{fig:util}
\end{figure}

\subsubsection{Real application performance}
We implemented a remote key-value store on \sys's testbed and used a cycle-accurate FPGA hardware simulator to evaluate \sys against the baselines using the YCSB workload~\cite{ycsb}.
\begin{figure}[t!]
    % \hfill
% \begin{subfigure}[b]{\linewidth}
    \centering
    \includegraphics[width=0.95\columnwidth]{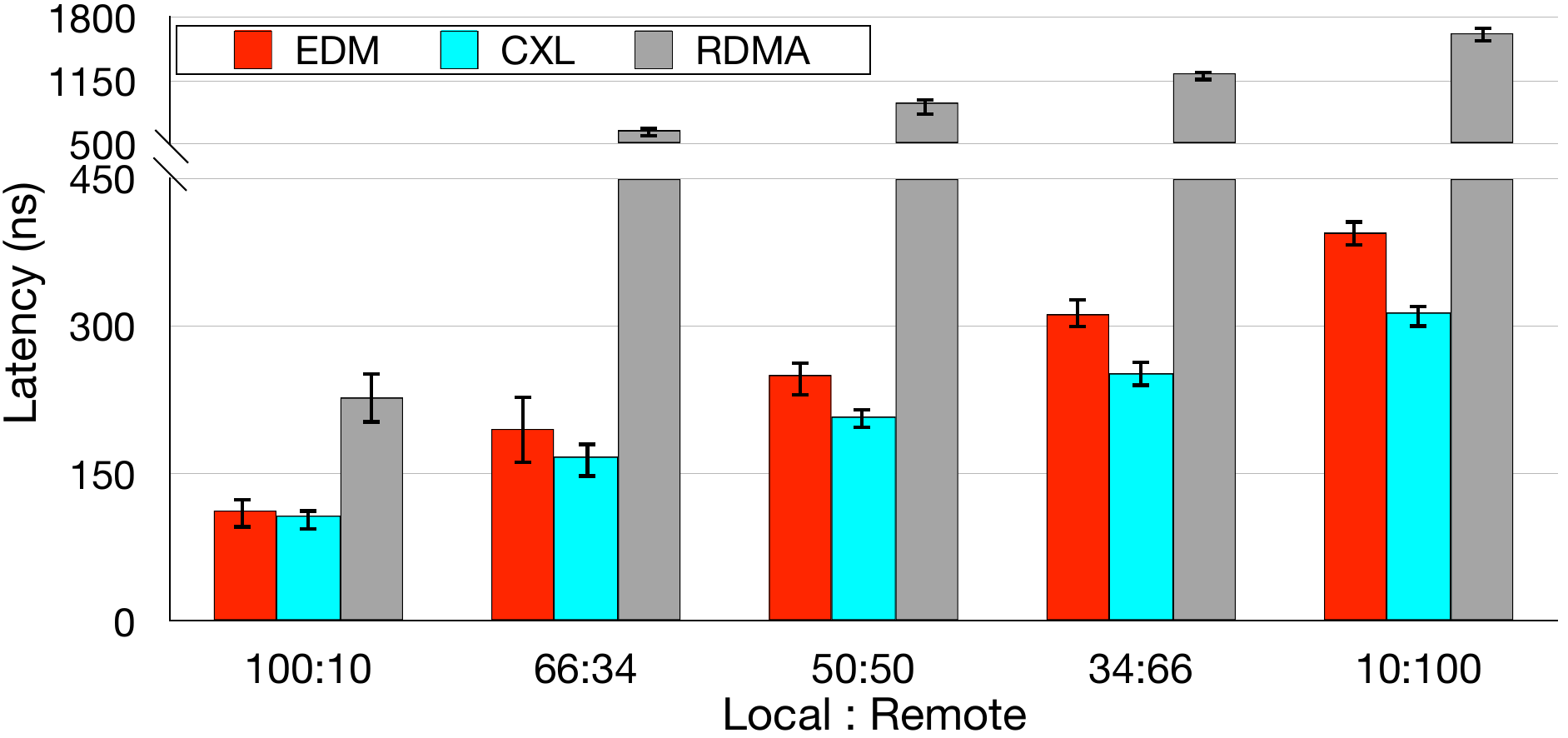}
    % \end{subfigure}
        \caption{End-to-end latency for YCSB workload.}
    \label{fig:e2e}
\end{figure}

\mypar{Bandwidth utilization} We stored the entire key-value store on the remote memory node and generated read/write requests at the compute node based on YCSB's workloads A, B and F~\cite{ycsb}. Each workload has a different read-write request distribution, A: 50\% write, B: 5\% write, and F: 33\% write. Each remote read request (8 B) queries for 1 KB data, while a remote write request carries 100 B data. \autoref{fig:util} shows that \sys is able to achieve around $2.7\times$ more throughput than RDMA in terms of requests per second. This is because of a combination of RoCEv2's transport protocol overhead, MAC layer minimum 64 B frame size constraint, and IFG overhead (\autoref{sec:background}). In contrast, in the PHY, \sys is able to avoid transport encapsulation, transmit data in 66-bit blocks, and repurpose the IFG to send memory traffic (\autoref{sec:design:stack}), thus saturating the entire link. Overall, this experiment highlights the benefit of operating in the PHY for the utilization of the link bandwidth. 
% Note that we do not batch requests here and that it 
% Note that in the above experiments, we could have improved the bandwidth utilization by batching multiple WREQ and RREQ flows into a single large frame. However, note that batching is only possible when multiple WREQ and RREQ flows are destined to the same destination. While we only have a single destination in our testbed, thus making batching a viable option, in practice, for larger networks with many destination memory nodes, the possibility of batching may be limited. Hence, this experiment shows the worst-case behavior of any protocol running on top of Ethernet MAC when no batching might be possible.

%Note that there is a decreased throughput of both in workload B because it has $95\%$ of reads which shared the link. Usually, the throughput of read requests is bottlenecked by responses (1KB) from memory node.

\mypar{Latency}
%In this experiment, we evaluate \sys's performance for a key-value store application using YCSB workload. To set up a memory-to-memory path, we direct-connect two FPGA boards, one as compute and the other as memory node. Note that the compute node also has a local DRAM module.
We distribute the key-value objects in different ratios between the local DRAM at the compute node and the remote DRAM at the memory node.
The compute node issues read/write requests based on YCSB's workload A to either the remote or its local memory.
We report the average request latency and variance in \autoref{fig:e2e}.
%However, YCSB trace pattern is Zipfian, for which the popular items' distribution will introduce variance to final result. Hence, for fairness, we randomly divide dataset into local (i.e., in compute node) and remote (i.e., on memory node) parts by different proportion and present average latency of multiple runs plus error bars.
We compare \sys's performance against CXL simulation~\cite{li2023pond} and RDMA measurement in \autoref{table:latency}. 
\autoref{fig:e2e} shows that \sys achieves a significantly lower latency than RDMA while within $1.3\times$ the latency of CXL. Note that although \sys is slightly slower than CXL in this small, unloaded testbed, in \autoref{sec:eval:simulations} we show that CXL scales much worse than \sys with a larger number of nodes and higher network load.
\begin{figure*}[h!]
    \centering
   \includegraphics[width=\linewidth]{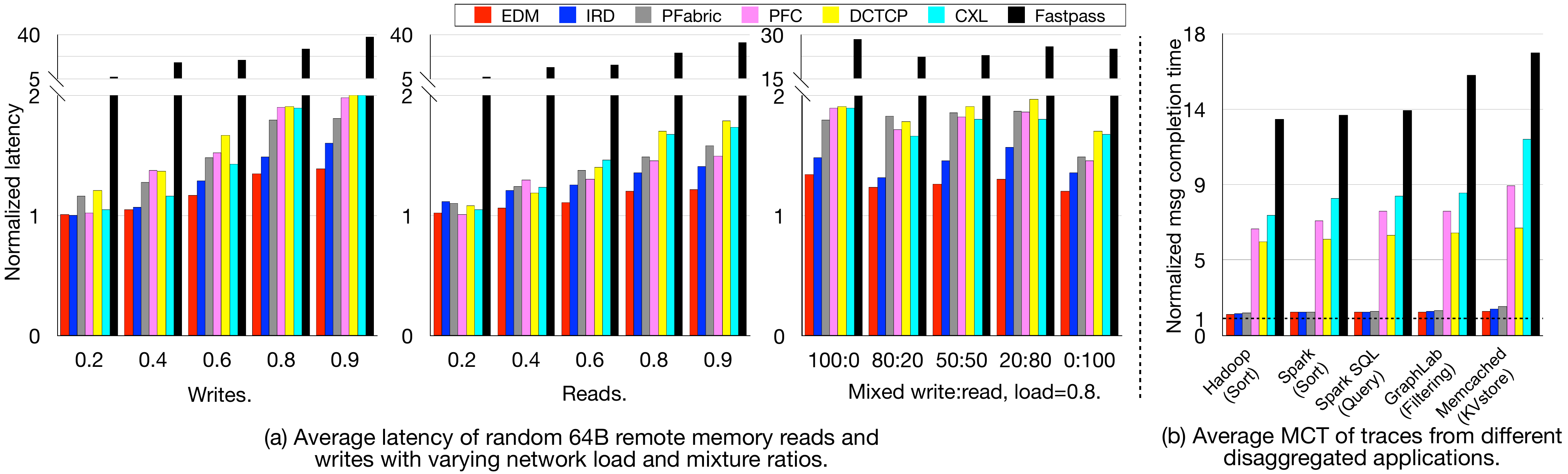}
    \caption{Network simulation.}
    \label{fig:netsim}
\end{figure*}
\subsection{Network Simulations}
\label{sec:eval:simulations}
To evaluate \sys's performance at scale, we complemented our hardware testbed with a network simulator written in C. The goal of the simulator is to evaluate the efficacy of \sys's in-network scheduler at scale under varying network loads.\smallskip\\
\noindent
\textbf{Set up.} We simulated a disaggregated cluster with 144 nodes (compute + memory) connected using a single switch. The nodes implement \sys's host stack, while the switch implements \sys's switch stack and the in-network scheduler. The latency of each of the \sys's modules and the propagation delay were configured to match the testbed from \autoref{table:latency}. We scaled the link bandwidth to 100\unit{Gbps}.\smallskip\\
\textbf{Baselines.} We evaluated \sys's scheduler against six classes of congestion and flow control protocols:
\textbf{(i) DCTCP}~\cite{dctcp}, a representative sender-driven protocol; \textbf{(ii) IRD}, an idealized receiver-driven protocol~\cite{ndp,phost,expresspass,homa,cai2022dcpim} that notifies each receiver of new flows in zero time. It combines the best features of existing protocols, like SRPT scheduling (from Homa~\cite{homa}, pHost~\cite{phost}, NDP~\cite{ndp}), and switch credit rate-limiting (from Expresspass~\cite{expresspass}); \textbf{(iii) pFabric}~\cite{pfabric}, a representative in-network scheduling protocol running on top of DCTCP; \textbf{(iv) PFC}~\cite{pfc}, a lossless flow control with DCQCN~\cite{dcqcn} as congestion control; \textbf{(v) CXL}, a lossless interconnect relying on PCIe's inherent credit-based flow control~\cite{cxl, aurelia}; and \textbf{(vi) Fastpass}~\cite{fastpass}, a centralized server-based flow scheduler, with 100 Gbps server bandwidth and an idealized assumption that it can solve the global scheduling problem infinitely fast. \smallskip \\
\textbf{\sys scheduler parameters.} We set the chunk size to 256\unit{B} and the maximum number of active notifications allowed per source--destination pair to 3.

\subsubsection{Latency microbenchmark}
\label{sec:eval:simulations:micro}
This experiment evaluated the effect of network load on remote memory access latency reported in \autoref{sec:eval:prototype:latency}. We generated all-to-all traffic between the compute and memory nodes with varying load values. The workload comprised a mixture of random remote memory reads and writes of size 64 B, with 8 B RREQ.

We calculated the latency of each read and write and normalized them by the corresponding unloaded latency (\autoref{table:latency}). As shown in \autoref{fig:netsim}a, \sys's read is within $1.2\times$, and write is within $1.3\times$ of its unloaded latency across all load values. IRD performs close to \sys at low load, but degrades quickly due to bandwidth under-utilization from its decentralized scheduling (\autoref{sec:background:sota}), which highlights the benefit of centralized scheduling.
Sender-driven protocols such as DCTCP, pFabric, and PFC perform poorly at high load due to their reactive nature (\autoref{sec:background:sota}), highlighting the benefit of proactive congestion control in \sys. Typically, pFabric outperforms DCTCP, but here their performance is identical due to uniformly single-packet flows in the workload, which makes SRPT scheduling used in pFabric ineffective. Finally, Fastpass uses a centralized scheduler similar to EDM, but the bandwidth of the central server becomes the bottleneck (which is more than 100$\times$ less than the aggregate cluster bandwidth), as it is overwhelmed by the large number of per-flow notification and grant messages. This highlights the benefit of in-network scheduling on switch hardware. Finally, CXL's point-to-point flow control fails in congested environment, as frequent incasts would rapidly consume credits on switch egress ports (victim) and that the deficit would consequently block or slow down all other ingress ports that have traffic destined to the victim. This is similar to head-of-line blocking in PFC~\cite{dcqcn}.

In addition, we also conducted an experiment that included a mixture of RREQ and WREQ in different ratios. \autoref{fig:netsim}a shows that \sys still achieves within 1.3$\times$ its unloaded latency for mixed requests. 
\subsubsection{Disaggregated application workloads}
\label{sec:eval:simulations:disaggregated}
Finally, we evaluated \sys on disaggregated memory workloads, based on public traces~\cite{shoal,osdi16:disaggregated}. 
% The traces comprise a variety of real-world applications, including batch processing, graph processing, interactive queries, and relational queries (see \autoref{sec:appendix:workloads} for a more detailed description of traces). 
The traces have a mixture of memory read and write requests in equal proportion with a heavy-tailed request size distribution that are derived from a variety of real-world applications as described below. 
\begin{itemize}[leftmargin=*]
\setlength\itemsep{0em}
\item \textbf{Hadoop (Sort).} Dataset: Sort benchmark generator.
\item \textbf{Spark (Sort).} Dataset: Sort benchmark generator.
\item \textbf{Spark SQL (Query).} Dataset: Big Data Benchmark~\cite{bdb}.
\item \textbf{GraphLab (Filtering).} Dataset: Netflix movie rating data.
\item \textbf{Memcached (Key-value Store).} Dataset: YCSB~\cite{ycsb}.
\end{itemize}
We measured the message completion time (MCT) for each memory message and normalized it by the ideal completion time, i.e., the time it would take for that message to complete if it were the only message in the network. \autoref{fig:netsim}b reports the average MCT for \sys is between 1.2--1.4$\times$ the ideal MCT. In contrast, MCTs for certain baseline protocols can be extremely high, in particular, MCT for CXL can be up to 8$\times$ higher than \sys, due to the reasons mentioned in \autoref{sec:eval:simulations:micro}. \\
%\textbf{Tail Performance.} The 99-percentile tail performance follows a similar trend as average. \sys consistently sustains low latency for both 64\unit{B} messages in \autoref{sec:eval:simulations:micro} (within $2\times$ unloaded latency) and real-world traces in \autoref{sec:eval:simulations:disaggregated} (within $3.5\times$).

\vspace{-0.5cm}
\section{Related Work}
%\noindent\textbf{Memory disaggregation.}

Conventional memory disaggregation architectures relied on RDMA for remote memory access, as seen in FaRM~\cite{dragojevic2014farm}, Infiniswap~\cite{gu2017efficient}, CFM~\cite{amaro2020can}, and Redy~\cite{zhang2021redy}. Unfortunately, the latency for remote memory access over RDMA can be of the order of $\mu$s, as shown in our evaluation.

Over the years, there have also been several proposals for novel hardware support for memory disaggregation~\cite{calciu2021rethinking,wang2020semeru,calciu2019project, novakovic2014scale, gouk2022direct,li2023pond,10018233}. 
Notably, some of these works propose a custom network fabric for better latency and cache-coherency.
For instance, soNUMA's~\cite{novakovic2014scale} fabric bears semblance to the Intel QPI and AMD HTX that interconnect sockets into multiple NUMA domains. Aquila~\cite{aquila} uses a custom cell-switched Dragonfly network fabric. Shoal~\cite{shoal} uses a fabric of ns-scale reconfigurable circuit switches with a focus on building a power-efficient rack-scale network for resource disaggregation. While, Compute Express Link (CXL)~\cite{gouk2022direct,li2023pond,10018233} facilitates low latency remote memory access through the PCIe link. In contrast, \sys is built upon Ethernet, which has several benefits over custom fabrics, as discussed in \autoref{sec:background:cxl}.

Recent years have also seen a growing interest in leveraging in-network computing for more efficient remote memory access~\cite{mind,chen2023cowbird,rpc}. These works advocate for offloading key memory management mechanisms, such as cache coherence and memory address translation, to a programmable switch.

\label{sec:related}
\vspace{-0.1cm}
\section{Conclusion}
\label{sec:conclusion}
We presented \sys, an ultra-low latency network fabric for memory disaggregation over Ethernet. \sys is built around two key design ideas. First, bypassing the Ethernet MAC layer and implementing the network protocol stack for remote memory access entirely in the Ethernet PHY, thus eliminating fundamental latency and bandwidth overheads imposed by the MAC layer; and second, running a centralized memory traffic scheduler in the switch's PHY to fully eliminate the latency due to queuing and layer 2 packet processing at the switch for memory traffic. Based on an FPGA hardware testbed, we show that \sys's fabric only incurs a latency of $\sim$300 ns in an unloaded network, for both remote memory reads and writes. Using larger-scale network simulations, we show that even at high network loads, \sys's average latency is within 1.3$\times$ its unloaded latency.
\vspace{-0.1cm}
\section*{Acknowledgments}
We would like to thank the anonymous reviewers and shepherd, Jichuan Chang, for their valuable feedback. This work was partially supported by NSF grants CAREER-2239829, CNS-2331111, and CCF-2402852.

\clearpage

\appendix

\section{Artifact Appendix}

%%%%%%%%%%%%%%%%%%%%%%%%%%%%%%%%%%%%%%%%%%%%%%%%%%%%%%%%%%%%%%%%%%%%%
\subsection{Abstract}

The artifact consists of three main components:
\begin{enumerate}
    \item A Verilog implementation of \sys PHY. This implementation demonstrates the practical feasibility and unloaded latency of \sys's design (\autoref{table:latency}).
    \item A cycle-accurate FPGA hardware simulator showing \sys's end-to-end performance in terms of latency and bandwidth utilization for a real-world application over realistic workloads (\autoref{fig:util} and \ref{fig:e2e}).
    \item A network simulator in C, demonstrating EDM's performance at scale and under varying network loads, particularly showcasing the efficacy of its in-network scheduling mechanisms (\autoref{fig:netsim}).
    
\end{enumerate}

Together, these components provide comprehensive cases supporting the two key technical contributions of our paper:
\begin{itemize}
    \item The implementation of host and switch network stacks in the Ethernet PHY for remote memory access.
    \item The design of a centralized, fast, in-network memory traffic scheduler in the PHY.
\end{itemize}

\subsection{Artifact check-list (meta-information)}

{\small
\begin{itemize}
  \item {\bf Program:}  FPGA project and C simulator.
  \item {\bf Compilation: } Vivado and GCC.
  \item {\bf Run-time environment: } Vivado 2023.2 on Ubuntu 22.04.3.
  \item {\bf Hardware: } Xilinx Alveo U200 FPGA board.
  \item {\bf Metrics: } Latency, bandwidth utilization, and message completion time.
  \item {\bf How much time is needed to complete experiments (approximately)?:} For hardware synthesis and experiments, it takes around two hours. Software network simulation experiments should finish within one hour.
  \item {\bf Publicly available?: } Yes. DOI: 10.5281/zenodo.14377804
  \item {\bf Code licenses (if publicly available)?: } MIT.
\end{itemize}
}

%%%%%%%%%%%%%%%%%%%%%%%%%%%%%%%%%%%%%%%%%%%%%%%%%%%%%%%%%%%%%%%%%%%%%
\subsection{Description}

\subsubsection{How to access}

\url{https://github.com/wegul/EDM}

\subsubsection{Hardware dependencies}

Xilinx Alveo U200 board.

\subsubsection{Software dependencies}

Vivado (validated on versions 2022.1 and 2023.2), Python>=3.9.

\subsubsection{Data sets}
\begin{enumerate}
    \item For hardware testbed experiments that focus on remote memory access latency, we use the read response and write request message sizes of 64 B, while the read requests are 8 B.
    \item For cycle-accurate FPGA hardware simulations running a real-world application, we use YCSB workloads~\cite{ycsb} with different read-write distributions.
    \item For software network simulations, we collected message traces from a diverse set of real-world disaggregated applications~\cite{osdi16:disaggregated} and analyzed the statistical distribution of message sizes to generate synthetic traces with different network loads.
\end{enumerate}

\subsection{Installation}

Hardware validation requires Vivado licensing; please follow \url{https://docs.amd.com/r/en-US/ug892-vivado-design-flows-overview/Design-Flows} and instructions in our repository.

For hardware and software network simulations, please follow the instructions in the repository on how to build and run the simulators.

\subsection{Experiment workflow}

\subsubsection{Hardware verification and simulation}
Our code comprises two components: custom modules integrated into the project library and modified native Ethernet modules with overloaded functionality. The full testbed consists of three Xilinx Alveo U200 FPGA boards, configured as compute, memory, and switch nodes, respectively. For convenience of artifact evaluation, we provide an example of a single board with loopback connection.

With latency metrics gathered from the hardware testbed, we perform cycle-accurate hardware simulation to obtain a more comprehensive evaluation of \sys's end-to-end performance for a real application over realistic workloads.

\subsubsection{Software Network simulation}
The simulation setup consists of two components: the trace generator and the network simulator. The trace generator leverages pre-existing CDF profiles of disaggregated workloads to generate experimental traces.
The network simulator incorporates seven distinct scheduling algorithms: EDM’s in-network scheduler and six state-of-the-art congestion and flow control mechanisms as comparative benchmarks. 

Our evaluation framework encompasses four distinct experimental scenarios: one utilizing real-world traces and three employing synthesized workloads.

\subsection{Evaluation and expected results}

%The artifact is structured to facilitate three primary validation objectives: hardware implementation verification of our PHY implementation, hardware simulation for end-to-end performance metrics, and evaluation of \sys central scheduler across diverse network parameters and traffic loads.

The hardware testbed component encompasses our complete implementation of the \sys host and switch network stack, allowing verification of the latency measurements presented in Table~\ref{table:latency}.

The hardware simulation experiments complement the hardware testbed with the YCSB workloads~\cite{ycsb} running over a key-value store application. It reproduces results in \autoref{fig:util} and \autoref{fig:e2e}.

For software network simulations, we provide a comprehensive automation framework consisting of scripted tools for trace generation and experimental execution, plus Python programs to visualize the comparative analysis to reproduce the results in \autoref{fig:netsim}.

%%%%%%%%%%%%%%%%%%%%%%%%%%%%%%%%%%%%%%%%%%%%%%%%%%%%
% When adding this appendix to your paper, 
% please remove below part
%%%%%%%%%%%%%%%%%%%%%%%%%%%%%%%%%%%%%%%%%%%%%%%%%%%%

\bibliographystyle{plain}
\balance
\bibliography{references}

\end{document}